\documentclass[]{aastex631}

\usepackage{amsmath}

\usepackage[normalem]{ulem}
\usepackage{enumitem}

\usepackage{newunicodechar,graphicx} 
\DeclareRobustCommand{\okina}{%
  \raisebox{\dimexpr\fontcharht\font`A-\height}{%
    \scalebox{0.8}{`}%
  }%
}
\newunicodechar{ʻ}{\okina}

\renewcommand{\leq}{\leqslant}  
  
\newcommand{\pow}[1]{^{#1}}
\newcommand{\poww}[1]{\pow{2}}
\newcommand{\powww}[1]{^{3}}

\newcommand{\pderiv}[2]{\partial_#2 #1}

\newcommand{\bs}[1]{\boldsymbol{#1}}
\newcommand{\mathbfit}[1]{\bs{\mathit{#1}}}
\renewcommand{\a}{{\mathbfit{a}}}
\renewcommand{\b}{{\mathbfit{b}}}
\newcommand{\e}{\hat{\bs{e}}}
\newcommand{\g}{\mathbfit{g}}
\renewcommand{\k}{\mathbfit{k}}

\newcommand{\m}{\mathbfit{m}}

\renewcommand{\v}{{\mathbfit{v}}}
\newcommand{\x}{\mathbfit{x}}
\newcommand{\B}{{\mathbfit{B}}}

\newcommand{\E}{{\mathbfit{E}}}
\newcommand{\F}{{{{}_2F_3}}}
\newcommand{\X}{\mathbfit{X}}
\renewcommand{\P}{{\mathbfit{P}}}
\newcommand{\U}{\mathbfit{U}}

\newcommand{\D}{\mbox{\boldmath$\mathsf{D}$}}
\newcommand{\R}{\mbox{\boldmath$\mathsf{R}$}}
\newcommand{\Q}{\mbox{\boldmath$\mathsf{Q}$}}
\newcommand{\J}{\mbox{\boldmath$\mathsf{J}$}}
\newcommand{\Y}{\mbox{\boldmath$\mathsf{Y}$}}

\newcommand{\vdot}{{\mathbf{\cdot}}}
\newcommand{\vcross}{{\mathbf{\times}}}
\newcommand{\grad}{\mbox{\boldmath$\nabla$}}

\newcommand{\diag}{\mathop{\rm diag}}

\newcommand{\curl}{\grad\vcross}

\renewcommand\div{\grad\vdot}

\newcommand{\alf}{Alfv{\'e}n}

\renewcommand{\a}{{\mathbfit{a}}}
\renewcommand{\b}{{\mathbfit{b}}}
\renewcommand{\k}{\mathbfit{k}}
\renewcommand{\v}{{\mathbfit{v}}}
\renewcommand{\P}{{\mathbfit{P}}}

\newcommand{\lrp}[1]{\left( #1 \right)}

\usepackage{xcolor}
\definecolor{forestgreen(web)}{rgb}{0.13, 0.55, 0.13}

\begin{document}

\title{Exact Nonlinear Decomposition of Ideal-MHD Waves Using Eigenenergies}

\author[0000-0002-6408-1829]{Abbas Raboonik}
\affiliation{The University of Newcastle,
University Dr, Callaghan,
NSW 2308, Australia}

\author[0000-0002-8259-8303]{Lucas A. Tarr}
\affiliation{National Solar Observatory, 22 Ohi\okina{}a Ku St, Makawao, HI, 96768}

\author[0000-0002-1089-9270]{David I. Pontin}
\affiliation{The University of Newcastle,
University Dr, Callaghan,
NSW 2308, Australia}



\begin{abstract}
In this paper we introduce a new method for exact decomposition of propagating, nonlinear magneto\-hydro\-dynamic (MHD) disturbances into their component eigenenergies associated with the familiar slow, Alfv\'en, and fast wave eigenmodes, and the entropy and field-divergence pseudo-eigenmodes. 
First the mathematical formalism is introduced, where it is illustrated how the ideal-MHD eigensystem can be used to construct a decomposition of the time variation of the total energy density into contributions from the eigenmodes. The decomposition method is then demonstrated by applying it to the output of three separate nonlinear MHD simulations.
The analysis of the simulations confirms that the component wave modes of a composite wavefield are uniquely identified by the method. The slow, Alfv\'en, and fast energy densities are shown to evolve in exactly the way expected from comparison with known linear solutions and nonlinear properties, including processes such as mode conversion. Along the way, some potential pitfalls for the numerical implementation of the decomposition method are identified and discussed.
We conclude that the exact, nonlinear decomposition method introduced is a powerful and promising tool for understanding the nature of the decomposition of MHD waves as well as analysing and interpreting the output of dynamic MHD simulations.
\end{abstract}



\section{Introduction}

The behaviour of plasmas in sufficiently low frequencies is modelled by the theory of magneto\-hydro\-dynamics (MHD). 
The MHD equations are hyperbolic and therefore support waves that transport energy through the system; Alfv\'en first noted this for the wave that now bears his name \citep{Alfven1942}, while the magneto\-acoustic waves were systematically studied by \citet{banos1955}.
The properties of these waves were rapidly worked out and can be found in any textbook on MHD or general plasma physics \citep[e.g.,][]{goedpoed04}.
MHD waves come in three basic modes, the \emph{slow}, \emph{Alfv\'en}, and \emph{fast} modes, each of which exhibits anisotropic propagation through the plasma.
The nature of that anisotropy and the nature of the waves themselves---how dominated their wave energy is by either magnetic or acoustic perturbations---is highly dependent on the local state of the plasma.
Accurately describing the anisotropic propagation of energy through an inhomogeneous plasma is the goal of the present work.

For certain plasma parameters, the eigenstructure of the MHD equations can become locally degenerate \citep{roe1996}, thereby allowing the local exchange of wave energy between the modes \citep{Cally2001}.
This has important implications for the way that energy is transported through a plasma, especially one like the solar atmosphere that is gravitationally stratified and highly magnetically-structured.
Identification and interpretation of wave modes and their importance for the dynamics are substantially complicated by both the variable nature of each type of wave and the tendency for energy exchange between modes (either by linear mode conversion or nonlinear interaction). 
This challenges our understanding of plasma dynamics in, e.g., observations of the solar atmosphere (as emphasized in the introduction of \citet{Cally2017}), the solar wind \citep[e.g.][]{ofman2010}, and in simulations \citep[see, e.g., the discussion in Section 4 of ][]{vandoorsselaere2020}.

In slightly more detail, waves are nonlocal phenomena, and their global description requires a complete specification of the plasma in a given volume along with boundary conditions.
For inhomogeneous plasmas and magnetic fields, the global eigenmodes do not neatly map onto the distinct slow, \alf{}, and fast modes of homogeneous MHD at all locations and for all times; however, at any given location and time, the global modes can still be classified as combinations of these underlying modes, a feature that can extend even beyond ideal MHD \citep[e.g., see][for a discussion of modes in Hall-MHD]{Raboonik2021}.
Fully analytic descriptions of the global modes are possible only for relatively simple systems, for instance, for a 2.5D linear gravitationally stratified atmosphere threaded by a uniform magnetic field \citep{Zhugzhda1984,Cally2001a}. 
Still, studies of these simplified systems demonstrate precisely how the character of a global mode will vary over space, and how, locally, that change in character is interpreted as mode conversion between the three underlying modes of the equivalent homogeneous system \citep{Cally2001a}.

To study energy propagation in  generic situations (lacking the symmetries required for the analytical solutions), a local --  rather than global -- approach can be employed. This is especially relavant for MHD simulations that focus on initial-boundary-value problems: for such problems, the goal is often to determine, for a certain type of energy injected into the system at Location A, what form the energy takes once (and if) it arrives at Location B.
Proposed methods to analyze the character of waves present in a simulation therefore typically take a local approach, both by necessity and by preference. That is the approach we take here.

In the context of the solar corona, where such inhomogeneities are universal, research into the behaviour of MHD waves has two principal motivations. The first is that damping of MHD waves by various mechanisms has long been invoked to explain the heating of various solar atmospheric structures \citep[e.g.][]{vandoorsselaere2020}, with the other being the utility of observations of such waves for deriving non-observable quantities through coronal seismology \citep[e.g.][]{demoortel2012,nakariakov2020}. In both applications it is highly desirable to be able to understand the characteristics of propagating disturbances in the context of simple, linear wave modes, whose properties are well studied.  Computational simulations addressing these problems are, by necessity, becoming increasingly sophisticated, and modern simulations can approach the complexity of the Sun itself in certain aspects, making interpretation of the resulting dynamics increasingly challenging.  
The correct characterisation of propagating disturbances in such simulations is therefore critical when comparing with observations. In the simplest approach, interesting dynamical behaviors are identified in simulations, and are interpreted in the context of linear MHD modes.
For example, \cite{wyper2022} simulated a bursty interchange reconnection event, identifying the production of Alfv\'enic waves in the open field region. Such waves are of great interest just now as they may form the seeds for the switchbacks observed in abundance in the solar wind in the latest observations \citep{bale2019}.  Many other studies also describe propagating disturbances produced by dynamic reconnection events \citep{mclaughlin2012,fuentes2012,wyper2014,stevenson2015,thurgood2017}. Each of these studies used approximate methods to characterise the waves in the system, relying on analogy to the properties and speeds of simple, linear MHD wave modes.

A more sophisticated approach involves a \emph{decomposition} of an identified propagating disturbance into slow, fast, and Alfv\'en(ic) contributions.
Most previous methods for decomposing the MHD wave field observed from numerical simulations rely on projecting the velocity field onto the directions parallel and perpendicular to the local magnetic field.
\citet{RosBogCar02aa} provide a prototypical example of this in 2D, while different methods have been used in 3D \citep[e.g.][]{2019A&A...625A.144R,2019ApJ...883..179K} as, reviewed in Section 3 of \citet{yadav2022} 
who go on to propose a new scheme of their own.
These methods boil down to interpreting the local properties of the velocity field by analogy to the closest simple MHD mode in the extreme limit of a  zero-$\beta$ or infinite-$\beta$ plasma.

\citet{Tarr2017} proposed a different decomposition approach where the dominance of the magnetic or acoustic terms in the local wave energy density was used to identify the different modes, i.e., fast (slow) waves in a low-$\beta$ environment have a magnetically (acoustically) dominated wave potential energy, and vice versa in a high-$\beta$ environment.
This approach utilizes the entire MHD state vector to identify waves instead of just the velocity components and more easily reveals the mixed-mode nature of the waves in regions of moderate $\beta$ that are difficult to interpret in velocity-only analyses.
The propagation of these wave energy densities could be directly tracked through the simulation over time, and then compared to the predicted speed and direction at which each type of wave energy density propagated through the simulation derived using ray tracing in the WKB limit \citep{weinberg1962}. 
This comparison helped to establish the validity of the decomposition and its interpretation \citep[see, e.g., chapter 2 of][for a detailed discussion on the WKB ray tracing approach]{mythesis}. \cite{Raboonik2019} use a similar approach in identifying the nature of linear MHD waves using the dispersion diagrams of linear 2.5D global modes that are specific to the state of the plasma.

Despite the utility of each of the above approaches in their individual applications, none of them are generally applicable because they only approximate the proper decomposition of plasma dynamics into the allowed motions of the plasma.
The local description of allowed displacements of a hyperbolic system can be generalized to the nonlinear regime via the method of characteristics; \citet{Jeffrey1964} provide a full exposition on the subject for MHD.  Indeed, such a description forms the basis of many numerical MHD methods \citep{HarLax83} and a (linear) version has been used to identify waves in solar wind observations \citep{Zank2023}.
Recently, \citet{Tarr2024} demonstrated how the characteristic description of MHD could be used to locally decompose and reconstruct the dynamical evolution of an MHD system in terms of these fully nonlinear characteristic modes.  While those authors focused on using the formalism to properly implement boundary conditions, they suggested that the method is underutilized for the \emph{analysis} of simulations, particularly for the unsolved problem of tracking the propagation and interaction of MHD waves in simulations.  The present study demonstrates the extension of the formalism and successful application toward that goal.

The paper is structured as follows. In \autoref{sec:math} we present the mathematical formalism of the eigenenergy decomposition method. The method is first illustrated in \autoref{sec:sim1D} through application to two 1D simulations, one with $\beta< 1$ and the other with $\beta> 1$. Then in \autoref{sec:sim3D} we apply the method to the propagation of disturbances in a fully 3D magnetic field. Finally, in \autoref{sec:conc} we present our conclusions.

\section{The eigenenergy Decomposition Method}\label{sec:math}
\subsection{The MHD Eigensystem}\label{sec:2p1}
The core idea behind the proposed decomposition method is the use of the ideal-MHD eigensystem to construct local transformations of the solutions onto the characteristic curves, where the fundamental MHD modes are discernible. We  stress the word ``solutions'' as throughout this paper it is assumed that the MHD solutions are fully known \textit{a priori}, whether by direct numerical simulation or other means, and here we merely use those solutions as input to analyse the MHD modes that exist therein.
Once the local transformation is known, we may break down the rate of change of the total energy density and ultimately tease out the energy contribution from each fundamental mode. For the sake of brevity, throughout this paper we will use the word energy to mean energy density.

The full set of ideal-MHD equations describing a polytropic plasma in the absence of gravity can be written in SI units by separating the temporal and spatial derivatives as follows,
\begin{subequations}\label{eq:3DidealMHD}
    \begin{equation}\label{eq:3DidealMHD:mass}
        \pderiv{\rho}{t} + \div{\lrp{\rho \bs{v}}} = 0,
    \end{equation}
    \begin{equation}\label{eq:3DidealMHD:momentum}
        \pderiv{\bs{v}}{t} + \bs{v}\vdot\bs{\nabla}\bs{v} + \frac{1}{\rho}\lrp{\grad{p} + \bs{j}\vcross\bs{B}} = 0,
    \end{equation}
    \begin{equation}\label{eq:3DidealMHD:induction}
        \pderiv{\bs{B}}{t} - \bs{B}\vdot\bs{\nabla}\bs{v} - \bs{B}\div{\bs{v}} - \bs{v}\vdot\bs{\nabla}\bs{B} = 0,
    \end{equation}
    \begin{equation}\label{eq:3DidealMHD:energy}
        \pderiv{p}{t} + \bs{v}\vdot\bs{\nabla} p + \gamma p \div{\bs{v}} = 0,
    \end{equation}
\end{subequations}
where the quantities have their usual meaning, i.e., $\rho$ is the mass density, $\bs{v}$ is the plasma velocity, $\bs{B}$ is the magnetic field, $\bs{j} = \curl{\bs{B}}/\mu_0$ is the current density, $\mu_0$ is the permeability of vacuum, $\gamma$ is the heat capacity ratio (taken to be $5/3$ throughout this study), and $p = \kappa \rho^{\gamma}$, with $\kappa$ assumed to be constant (indicating an isentropic process). Note that the precondition on $\bs{B}$ to remain divergence-free is built into the induction equation \eqref{eq:3DidealMHD:induction}, and hence if $\div{\bs{B}} = 0$ initially, then it shall so remain throughout.

The system of equations above is hyperbolic, and hence can be recast into the quasi-linear form as follows \citep{roe1996}
\begin{equation}\label{eq:quasiLinear}
    \pderiv{\bs{P}}{t} + M_q\pderiv{}{q}\bs{P} = 0,
\end{equation}
in which $\bs{P} = \lrp{\rho, v_x, v_y, v_z, B_x, B_y, B_z, p}^{T}$ is the primitive plasma state vector (the solution vector mentioned above which is assumed to be known at all times), $M_q$ are $8\times8$ time- and space-dependent flux matrices in the direction $q$ (for which, from here on out, we assume $q \in \lrp{x,y,z}$, unless stated otherwise), and summation is implied over the repeated indices. Without imposing any simplifying assumption, the flux matrices in 3D Cartesian coordinates are found to be
\begin{subequations}\label{eq:fluxMatrices}
    \begin{equation}
    M_x(\bs{x},t) = \begin{pmatrix}
        v_x & \rho & 0 & 0 & 0 & 0 & 0 & 0\\
        0   & v_x  & 0 & 0 & 0 & B_y/\mu_0\rho & B_z/\mu_0\rho & 1/\rho \\
        0 & 0 & v_x & 0 & 0 & -B_x/\mu_0\rho & 0 & 0 \\
        0 & 0 & 0 & v_x & 0 & 0 & -B_x/\mu_0\rho & 0 \\
        0 & 0 & 0 & 0 & v_x & 0 & 0 & 0 \\
        0 & B_y & -B_x & 0 & 0 & v_x & 0 & 0 \\
        0 & B_z & 0 & -B_x & 0 & 0 & v_x & 0 \\
        0 & \gamma p & 0 & 0 & 0 & 0 & 0 & v_x
    \end{pmatrix},
\end{equation}

\begin{equation}
    M_y(\bs{x},t) = \begin{pmatrix}
        v_y & 0 & \rho & 0 & 0 & 0 & 0 & 0\\
        0   & v_y  & 0 & 0 & -B_y/\mu_0\rho & 0 & 0 & 0 \\
        0 & 0 & v_y & 0 & B_x/\mu_0\rho & 0 & B_z/\mu_0\rho & 1/\rho \\
        0 & 0 & 0 & v_y & 0 & 0 & -B_y/\mu_0\rho & 0 \\
        0 & -B_y & B_x & 0 & v_y & 0 & 0 & 0 \\
        0 & 0 & 0 & 0 & 0 & v_y & 0 & 0 \\
        0 & 0 & B_z & -B_y & 0 & 0 & v_y & 0 \\
        0 & 0 & \gamma p & 0 & 0 & 0 & 0 & v_y
    \end{pmatrix},
\end{equation}

\begin{equation}
    M_z(\bs{x},t) = \begin{pmatrix}
        v_z & 0 & 0 & \rho  & 0 & 0 & 0 & 0 \\
 0 & v_z & 0 & 0 & -B_z/\mu_0\rho & 0 & 0 & 0 \\
 0 & 0 & v_z & 0 & 0 & -B_z/\mu_0\rho & 0 & 0 \\
 0 & 0 & 0 & v_z & B_x/\mu_0\rho & B_y/\mu_0\rho & 0 & 1/\rho \\
 0 & -B_z & 0 & B_x & v_z & 0 & 0 & 0 \\
 0 & 0 & -B_z & B_y & 0 & v_z & 0 & 0 \\
 0 & 0 & 0 & 0 & 0 & 0 & v_z & 0 \\
 0 & 0 & 0 & \gamma  p & 0 & 0 & 0 & v_z \\
    \end{pmatrix}.
\end{equation}
\end{subequations}

Given the above-mentioned hyperbolicity of the equations, the eigensystem associated with these matrices is analytical everywhere, and may be used to extract the required transformation matrices.\footnote{Note that the MHD equations are not strictly hyperbolic, as the eigensystem becomes degenerate in certain limits.  However, the system remains well behaved in those limiting cases for judicious choice of normalization \citep{brio1988, roe1996}.}
To this end, solving the eigenvalue problem $M_q R_q = R_q \Lambda_q$ (or similarly $L_q M_q = \Lambda_q L_q$), wherein $R_q$ ($L_q$) is the right (left) eigenmatrix, we find the eigenvalues to be
\begin{equation}\label{eq:evals}
    \Lambda_q = \diag{\lrp{v_q, v_q, v_q - a_q, v_q + a_q, v_q - c_{\text{s},q}, v_q + c_{\text{s},q}, v_q - c_{\text{f},q}, v_q + c_{\text{f},q}}},
\end{equation}
where $a_q = B_q/\sqrt{\mu_0 \rho}$ is the familiar $q$-directed \alf{} speed,
and $c_{\text{s},q}$ and $c_{\text{f},q}$ are the slow and fast speeds, respectively, given by
\begin{equation}
        c_{\text{f/s},q} = \frac{\sqrt{a^2+c^2\pm\sqrt{a^4+c^4+c^2 \left(a^2-2 a_q^2\right)}}}{\sqrt{2}},
\end{equation}
and $c = \sqrt{\gamma p/\rho}$ is the (isotropic) adiabatic sound speed. The corresponding right and left eigenmatrices ($R_q$ and $L_q$) are provided in \autoref{sec:AppendixA}.

Each eigenvalue in \autoref{eq:evals} pertains to a unique mode of the system of \autoref{eq:3DidealMHD}, and prescribes the group velocity of disturbances along the direction $q$. Therefore, given the convective nature of the first two eigenvalues (as implied by their respective group velocities being equal to the plasma velocity), they must correspond to (zero-frequency) pseudo-modes. This means that, so long as the plasma behaves precisely according to the set of \autoref{eq:3DidealMHD}, they do not propagate any energy, and are merely there to ensure that the zero magnetic field divergence requirement, as well as the plasma properties defining the entropy (which is constant under our polytropic assumption) are identically preserved and convectively transported throughout the system. 
For reasons that will become more evident from the mathematics, we shall term these to be the \textit{field divergence} and \textit{entropy} pseudo-modes.
On the other hand, as signified by the propagation velocities $\pm a_q$, $\pm c_{\text{s},q}$, and $\pm c_{\text{f},q}$ shifted relative to the bulk velocity $v_q$, the remaining six eigenvalues belong to the \alf, slow, and fast modes, respectively, with the $+$ and $-$ signs indicating the direction of propagation on the corresponding characteristic curve. Note that the characteristic curves for the \alf{} waves are simply the magnetic field lines, while the fast and slow waves can, in general, propagate obliquely. 

\subsection{Decomposition of the Total Energy}
We are now in a position to derive the mode-decomposed expression for the total energy $E_{\rm{tot}}$.
For a non-gravitational ideal plasma, $E_{\rm{tot}}$ is made out of kinetic (kin), internal (int), and magnetic (mag) parts given by
\begin{align}\label{eq:EnTot}
    E_{\rm{tot}} &= E_{\rm kin} + E_{\rm int} + E_{\rm mag} \nonumber \\
    &= \frac{1}{2} \rho v^2 + \frac{p}{\gamma - 1} + \frac{1}{2 \mu_0} B^2.
\end{align}
Differentiating \autoref{eq:EnTot} with respect to time yields
\begin{equation}\label{eq:ddtentot}
    \partial_t{E_{\rm{tot}}} = \frac{1}{2} \rho_t v^2 + \rho \bs{v}_t\vdot\bs{v} + \frac{p_t}{\gamma - 1} + \frac{1}{\mu_0} \bs{B}_t\vdot\bs{B}.
\end{equation}
On the other hand, using the relation $M_q = R_q \Lambda_q L_q$, we may rewrite \autoref{eq:quasiLinear} in the following format
\begin{equation}\label{eq:LLDV}
    \partial_t{\bs{P}} + R_q \bs{\mathcal{L}}_q(\bs{x},\bs{\nabla},t) =0,
\end{equation}
where $\bs{\mathcal{L}}_q(\bs{x},\bs{\nabla},t) = L_q \Lambda _q \partial_q{\bs{P}}$ is the vector of characteristic derivatives in the direction $q$. The $\bs{\nabla}$ argument means that $\bs{\mathcal{L}}_q$ contain spatial derivatives, while it is devoid of time derivatives. Therefore, \autoref{eq:LLDV} provides an important connection between the local rate of change of the state vector and the mode-decomposed quantity $R_q \bs{\mathcal{L}}_q$ that solely depends upon local spatial variations. It is the second term that encapsulates the physics of the system, and hence prescribes appropriate local transformations to be utilized in order to break down the total energy.

\autoref{eq:LLDV} provides a method to decompose the rate of change of each of the state variables in terms of each of the eight modes of the system. Thus,
substituting the time derivatives obtained from \autoref{eq:LLDV} into the RHS of \autoref{eq:ddtentot}, we may rewrite the rate of change of the total energy in terms of an 8-sum composed of terms representing the contribution of each mode:
\begin{align}\label{eq:ddtdecomposed1}
    \partial_t{E_{\rm{tot}}} &= \sum_{n = 1}^{8} \boldsymbol{c}_n(\bs{x},t) \boldsymbol{.}\overline{\mathcal{L}}_n(\bs{x},\bs{\nabla},t) \nonumber \\
    &= \sum_{q \in (x,y,z)}\lrp{\partial_t{E_{\text{div},q}} + \partial_t{E_{\text{ent},q}} + \partial_t{E_{\text{A},q}^{-}} + \partial_t{E_{\text{A},q}^{+}} + \partial_t{E_{\text{s},q}^{-}} + \partial_t{E_{\text{s},q}^{+}} + \partial_t{E_{\text{f},q}^{-}} + \partial_t{E_{\text{f},q}^{+}}},
\end{align}
in which $\boldsymbol{c}_n$ are the local vectors of proportionality formed by combining the coefficients of the time derivative terms in \autoref{eq:ddtentot} with elements of the $R_q$ matrices, and $\overline{\mathcal{L}}_n$ are ($3\times8$) matrices whose columns are made out of the $\bs{\mathcal{L}}_q$ vectors, with $n$ representing the mode numbers according to: ($n = 1, 2$) the field divergence (div) and entropy (ent) pseudo-modes, ($n = 3, 4$) the forward ($+$) and reverse ($-$) \alf{} (A) modes, ($n = 5, 6$) the forward and reverse slow (s) modes, and ($n = 7, 8$) the forward and reverse fast (f) modes, respectively, where forward and reverse refer to the two directions along each characteristic curve. The second line of \autoref{eq:ddtdecomposed1} shows the mode-expanded summation in the above-mentioned order, with this summation being over the directions $q$ instead.

The individual terms in the summation above, henceforth referred to as the eigenenergy time derivatives, are given by
\begin{subequations}\label{eq:EDT}
    \begin{equation}\label{eq:EDivdt}
        \partial_t{E_{\text{div},q}} = -\frac{B_q \partial_q{B_q}}{\mu_0} v_q,
    \end{equation}
    
    \begin{equation}\label{eq:EEntdt}
        \partial_t{E_{\text{ent},q}} = \frac{v^2 \lrp{\partial_q{p}- c^2 \partial_q{\rho}}}{2 c^2} v_q,
    \end{equation}

    \begin{equation}\label{eq:EAlfdt}
        \partial_t{E_{\text{A},q}^{\mp}} = \pm \frac{\rho \lrp{\pm \bs{v}\bs{\times}\bs{B}}_q \lrp{\lrp{\pm\partial_q{\bs{v}}\bs{\times}\bs{B}} + \lrp{\partial_q{\bs{B}}\bs{\times}\bs{a}}}_q}{2 B_\perp^2}\lrp{a_q \mp v_q},
    \end{equation}

    \begin{align}\label{eq:ESlowdt}
        \partial_t{E_{\text{s},q}^{\mp}} = \frac{\pm1}{8 c^2 c_{\text{s},q} \lrp{c_{\text{f},q}^2 - c_{\text{s},q}^2} \sqrt{\mu_0 \rho}}\lrp{\frac{2 c_{\text{s},q}^2}{a_q^2 - c_{\text{s},q}^2}\bs{a}_\perp\vdot\lrp{\bs{a}_\perp \pm \frac{a_q}{c_{\text{s},q}}\bs{v}_\perp} - v^2 \pm 2 c_{\text{s},q} v_q - \frac{2 c^2}{\gamma - 1}}\nonumber \\ \lrp{2 \rho c^2 \lrp{\pm B_q \bs{a}\vdot\partial_q{\bs{v}} + c_{\text{s},q} \bs{a}_\perp\vdot\partial_q{\bs{B}_\perp} \mp \frac{c_{\text{s},q}^2}{a_q} B_q \partial_q{v_q}}+\sqrt{\mu_0 \rho}\, c_{\text{s},q}\, \partial_q{p}\lrp{\lrp{c^2 - a^2} - \lrp{c_{\text{f},q}^2 - c_{\text{s},q}^2}}}\lrp{c_{\text{s},q} \mp v_q},
    \end{align}

    \begin{align}\label{eq:EFastdt}
        \partial_t{E_{\text{f},q}^{\mp}} = \frac{\pm1}{8 c^2 c_{\text{f},q} \lrp{c_{\text{s},q}^2 - c_{\text{f},q}^2}\sqrt{\mu_0 \rho}}\lrp{\frac{2 c_{\text{f},q}^2 }{a_q^2 - c_{\text{f},q}^2}\bs{a}_\perp\vdot\lrp{\bs{a}_\perp \pm \frac{a_q}{c_{\text{f},q}}\bs{v}_\perp} - v^2 \pm 2 c_{\text{f},q} v_q - \frac{2 c^2}{\gamma - 1}}\nonumber \\ \lrp{2 \rho c^2 \lrp{\pm B_q \bs{a}\vdot\partial_q{\bs{v}} + c_{\text{f},q} \bs{a}_\perp\vdot\partial_q{\bs{B}_\perp} \mp \frac{c_{\text{f},q}^2}{a_q} B_q \partial_q{v_q}}+\sqrt{\mu_0 \rho}\, c_{\text{f},q}\, \partial_q{p}\lrp{\lrp{c^2 - a^2} - \lrp{c_{\text{s},q}^2 - c_{\text{f},q}^2}}}\lrp{c_{\text{f},q} \mp v_q},
    \end{align}
\end{subequations}
in which $\bs{a}$ is the \alf{} speed vector, and the subscript $\perp$ denotes 2-vectors comprised of the components along the axes transverse to the direction $q$ according to the familiar Cartesian right-hand permutations (i.e., if $q = y$, then $\bs{q}_\perp = \lrp{z,x}$). 
We emphasize that \autoref{eq:EDT} are exact, nonlinear, and analytical everywhere in an ideal plasma. Note the entirely magnetic nature of \autoref{eq:EDivdt} and \ref{eq:EAlfdt} associated with the field-divergence pseudo-mode and the \alf{} mode, and the fully acoustic nature of \autoref{eq:EEntdt} corresponding to the entropy pseudo-mode, as one would expect. Moreover, when summed over $q$ and time-integrated, \autoref{eq:EDivdt} provides the energy version of the constraint on the magnetic field to ensure that it remains divergence free, while \autoref{eq:EEntdt} provides such a constraint on our assumed thermodynamical nature of the gas to remain polytropic. Of course, for numerical simulations, such constraints are not always precisely met at all times and locations and there will be deviations, due to resistivity, shock heating, etc.. As we will later see, this will cause some amounts of energy to be carried by the pseudo-modes, despite our analytically deduced statement that they should not contribute to the overall energy transport under the strictly exact ideal-MHD framework.

Note the relationship between the $\pm$ signs in \autoref{eq:EAlfdt} through to \ref{eq:EFastdt}, where the first two $\pm$ signs on the RHS of \autoref{eq:EAlfdt} are redundant, but purposely placed for clarity and consistency with the two subsequent equations. In all these cases, the reverse ($-$, $n = 3, 5, 7$) and forward ($+$, $n = 4, 6, 8$) expressions are simply connected by flipping the sign of the entire equation in addition to flipping the sign of all the velocity components and their derivatives on the RHS. This further provides insight on the differences between the forward and reverse modes, indicating that they indeed describe the direction of the flow of energy along the characteristics.

In light of \autoref{eq:EDT}, it is noteworthy that \autoref{eq:ddtdecomposed1} can also be recast as follows
\begin{equation}
    \partial_t{E_{\rm{tot}}} = \sum_{q \in (x,y,z)} \bs{\lambda}_q\vdot\bs{F}_q,
\end{equation}
where $\bs{\lambda}_q$ and $\bs{F}_q$ are 8-vectors comprised, respectively, out of the eight eigenvalues in each direction and their corresponding coefficients found in \autoref{eq:EDT}. The entries of $\bs{F}_q$ will all then share the same dimension of energy density per unit length, or simply force density.

The assumption that the solutions are known \textit{a priori} means that the quantities in \autoref{eq:ddtdecomposed1} are also known and quantifiable everywhere in the plasma. Thus, we postulate that the total energy itself can be broken down in terms of the individual energies carried by each ideal-MHD mode, derived by performing time integration of each of the terms in that equation according to
\begin{align}\label{eq:totendecomposed}
    E_{\rm{tot}} &= E_{\rm{rec}} + f(x,y,z) \nonumber \\ 
    &= \sum_{q \in (x,y,z)}\lrp{E_{\text{div},q} + E_{\text{ent},q} + E_{\text{A},q}^{-} + E_{\text{A},q}^{+} + E_{\text{s},q}^{-} + E_{\text,q}^{+} + E_{\text,q}^{-} + E_{\text{f},q}^{+}} + f(x,y,z),
\end{align}
where $E_{\rm{rec}}$ (given by the terms in the summation's brackets) is the net amount of pure propagating wave-energy of the dynamic ideal plasma as \textit{recovered} by the decomposition method, and $f$ is the function of integration corresponding to the static background energy. \autoref{eq:totendecomposed}, together with \autoref{eq:EDT} form the core of the eigenenergy decomposition method (EEDM).

As a first sanity test of the decomposition method formulated here, we refer the reader to \autoref{sec:AppendixB}, where we consider pure magneto\-acoustic waves, and illustrate that the method correctly attributes no energy to the \alf{} branch.

\subsection{Comments on Implementation}

\autoref{eq:totendecomposed} gives an expression for the total energy in terms of the component wave energies. Analytically, this should match the total energy as calculated from primitive code variables (Eq.~\ref{eq:EnTot}). However, this may not be the case in  a numerical implementation such as the one described below (due to numerical differentiation and integration in the decomposition method, see below and \autoref{sec:AppendixE}). In order to assess the fidelity of the decomposition we write the energy obtained from primitive code variables (Eq.~\ref{eq:EnTot}) as $E_{\text{tot}} = E_{\text{orig}} + E_{0}$, with $E_0$ being the energy of the static initial (or background) state, and $E_{\text{orig}}$ being the (``original'') energy of the fluctuating (wave) component -- obtained by subtracting the total energy from the initial total energy, using primitive code variables. When presenting our results in the subsequent sections, we will test the numerical fidelity of the decomposition method by providing comparisons between the net wave energies pre- and post-decomposition given by $E_{\rm orig}$ and $E_{\rm rec}$, respectively. 

Due to the mentioned lack of 3D nonlinear solutions, and in order to provide more insight into the method, we will have to resort to direct numerical simulations that are representative of sufficiently simple conditions to be able to interpret and substantiate the results. In the two following sections, we will discuss two ideal-MHD simulation setups along with their results. These are made of a uniform plasma superimposed by (I) a vertical magnetic field and excited by a circularly polarized driver aimed at primarily generating upward travelling \alf{} waves, and for comparison with previous work, (II) the 3D structured magnetic field of \cite{Galsgaard2003} containing a null point at the center and excited by the circular driver described therein, again aimed at maximising the generation of \alf{} waves. Due to the core differences between the two simulation setups, we will fully describe each case and discuss its results individually in the two separate sections that follow.

The mentioned drivers are introduced as boundary conditions (BCs), and once again we remind the reader that these do not constitute exact ideal-MHD solutions (but rather numerical ones), and hence some discrepancy is expected with the theoretical exposition above (explored below). This means that \alf{} waves will not be the only product of such drivers.
Note that the purpose of the simulations introduced here is
to verify the capacity of the decomposition method to identify wave modes, rather than to model any specific physical process in the solar atmosphere or elsewhere.

The numerical package used throughout this work is Lare3d \citep{ARBER2001151}, which is second-order accurate in space and time. The code is built on a staggered grid, and at each step advances the solutions to the dimensionless Lagrangian 3D MHD equations and then conservatively remaps them back onto the original grid using volume averaging techniques. So far we introduced and formulated the method using the dimensional MHD equations for the sake of more clarity. However, in view of the dimensionless nature of the Lare3d solver used for our simulations, for the remainder of this study we shall work with non-dimensional units. This is simply done by setting $\mu_0=1$ throughout our equations and treating all the other variables as non-dimensional.

\section{Simulation I: Uniform vertical magnetic field}\label{sec:sim1D}
In this section, we will expand on and examine the results of the first simulation (I) touched on before. This is a first step into empirical testing of the proposed decomposition scheme, aiming to compare its results with previously known asymptotic characteristics of ``linear'' MHD waves away from the equipartition level (at which $c = a$ and $\beta\approx 1$).  To this end, we divide simulation I into two categories of high-$\beta$ (hereafter referred to as simulation Ia) and low-$\beta$ (hereafter referred to as simulation Ib) and discuss both in detail.

The simulation box spans $-4.7 \leq x,y \leq 4.7$ in the horizontal plane and $0 \leq z \leq 32.92$ along the vertical axis. These dimensions are so chosen, first, to provide sufficient grid points for accurate numerical evaluations of the spatial derivatives present in \autoref{eq:EDT}, and second, to ensure a sufficiently large box for the wavelength of the driver (to be discussed shortly) and the group velocities of the waves it generates for data analysis purposes. The grid used to resolve the simulation contains 192$\times$192$\times$672 points, which yields resolution of $dx = dy = dz = 0.049$. The simulations are then run for (Ia) 40 and (Ib) 4 dimensionless time units, and snapshots are saved at a rate of (Ia) $dt = 0.05$, and (Ib) $dt = 0.01$. Such discretization schemes provide economically adequate spatial and temporal resolution for testing the \emph{post facto} decomposition method. To non-dimensionalize the MHD equations, we use the following scaling factors: $L_N = 150$ km, $\rho_N = 3.03 \times 10^{-4}$ kg.m$^{-3}$, $T_N = 5778$ K, respectively specifying the normalizing length, density, and temperature.

\subsection{Magneto\-hydro\-static model (initial conditions)}

For the initial state we consider a uniform plasma threaded by a uniform magnetic field of non-dimensional strength $B_0=0.1$ in the $z$-direction. In simulation Ia, the uniform density and temperature in the initial state are chosen as $\rho_0 = 0.0198$, and $T_0 = 2.2748$, giving a plasma-$\beta$ of $9.03$. In Ib, $\rho_0 = 2.4614 \times 10^{-5}$, $T_0 = 111.21$, and $\beta = 0.54$.
The remaining initial conditions on the pressure and internal energy are then easily found using the ideal gas equation. Lastly, to ensure the system is initially at rest, we set $\bs{v}_0 = 0$. We note that these parameter combinations are chosen to match two different heights from a plane-stratified model of the solar atmosphere (to be considered in a future paper).

\subsection{Driver and BCs: travelling torsional Alfv\'enic waves}
Given the symmetry of the model, we set the horizontal BCs to periodic, while choosing fully symmetric boundaries at the top plane. 
For the bottom boundary plane, we wish to set up simple torsional Alfv\'en conditions that incompressively excite travelling waves at $z = 0$. The goal is to primarily generate travelling \alf{} waves.
However, as there are no general exact nonlinear solutions to the ideal-MHD \autoref{eq:3DidealMHD} that would guarantee pure \alf{} waves (except for the semi-trivial case discussed in \autoref{sec:AppendixD}), we may seek approximate linear solutions of this kind, with the caveat that when used in the nonlinear regime, they will excite other waves as well. 

\cite{Turkmani2004} present 1.5\-D WKB solutions describing the approximate linear behaviour of ``circularly polarized'' torsional \alf{} waves in a gravitationally stratified isothermal plasma superimposed with a uniform vertical magnetic field $\bs{B} = B_0 \hat{\bf z}$.  For the purpose of this study under the uniform plasma assumption, we shall do away with the stratification and modify their solutions accordingly as follows
\begin{subequations}\label{eq:turkmani2004}
    \begin{equation}
        \bs{B}_\perp(z,t) = B_\perp(z,t)\lrp{\cos\lrp{k z - \omega t + \varphi},\sin\lrp{k z - \omega t + \varphi}},
    \end{equation}
    \begin{equation}\label{eq:vperp}
        \bs{v}_\perp(z,t) = -\frac{\bs{B}_\perp(z,t)}{B_0} a_0,
    \end{equation}
    \begin{equation}
        \omega = a_0 k,
    \end{equation}
\end{subequations}
where $a_0 = B_0/\sqrt{\rho_0}$ is the initial vertical component of the \alf{} speed, $k = 2 \pi/5$ is the wavenumber, $\omega$ is the \alf{} frequency, $\varphi = \pi/4$ sets the phase, and the subscript $\perp$ represents the directions perpendicular to the direction of propagation, i.e., the $z$-axis. 
To configure the driver to produce travelling waves moving initially at the \alf{} speed, we may localise the amplitude of the field fluctuations in space and time by choosing
\begin{equation}\label{eq:travellingAlf:Bperp}
    B_\perp(z,t) = A \exp{\lrp{-\frac{(z - a_0 t - z_0)^2}{2 \sigma^2}}},
\end{equation}
in which $A$ is a constant (the maximum amplitude, set to $2 \times 10^{-4}$ in simulation Ia and $10^{-2}$ in simulation Ib), $z_0 = -4.1556$ is the initial location of the pulse for a smooth introduction of the disturbances into the main domain, and $\sigma = 0.5$ sets the width of the pulse. 
Finally, setting $v_z = 0$ completes setting up our desired torsional \alf ic driver in the ghost cells. It should be noted that the described driver does not satisfy the $z$-component of the full momentum equation in the main simulation domain, which means that a component $v_z$ is generated in the first internal grid cell to balance this deficiency---see \autoref{sec:AppendixC} for a further discussion.

Before proceeding with presenting the simulation results, we note that the difficulties in finding an exact solution to the nonlinear MHD equations arise from the space and time localization of the \alf{} wave.
In \autoref{sec:AppendixD} we show that a monochromatic \alf{} wave, constructed by taking the limit of \autoref{eq:travellingAlf:Bperp} as $\sigma \to \infty$, does produce a semi-trivial solution of the MHD equations.
This is equivalent to setting $B_\perp$ to a constant.
The eigenenergy decomposition method can be analytically verified in that case, thus providing another sanity check on the validity of the method.

\subsection{Results}
We shall now individually present the results of the first family of simulations (Ia and Ib). The general formats in which the figures are presented are standardized to remain the same across this and the next section, for better tractability and readability, unless stated otherwise.

\subsubsection{Simulation Ia: high-\texorpdfstring{$\beta$}{beta}}
\begin{figure}
    \centering    \includegraphics[width=\textwidth,height=\textheight,keepaspectratio]{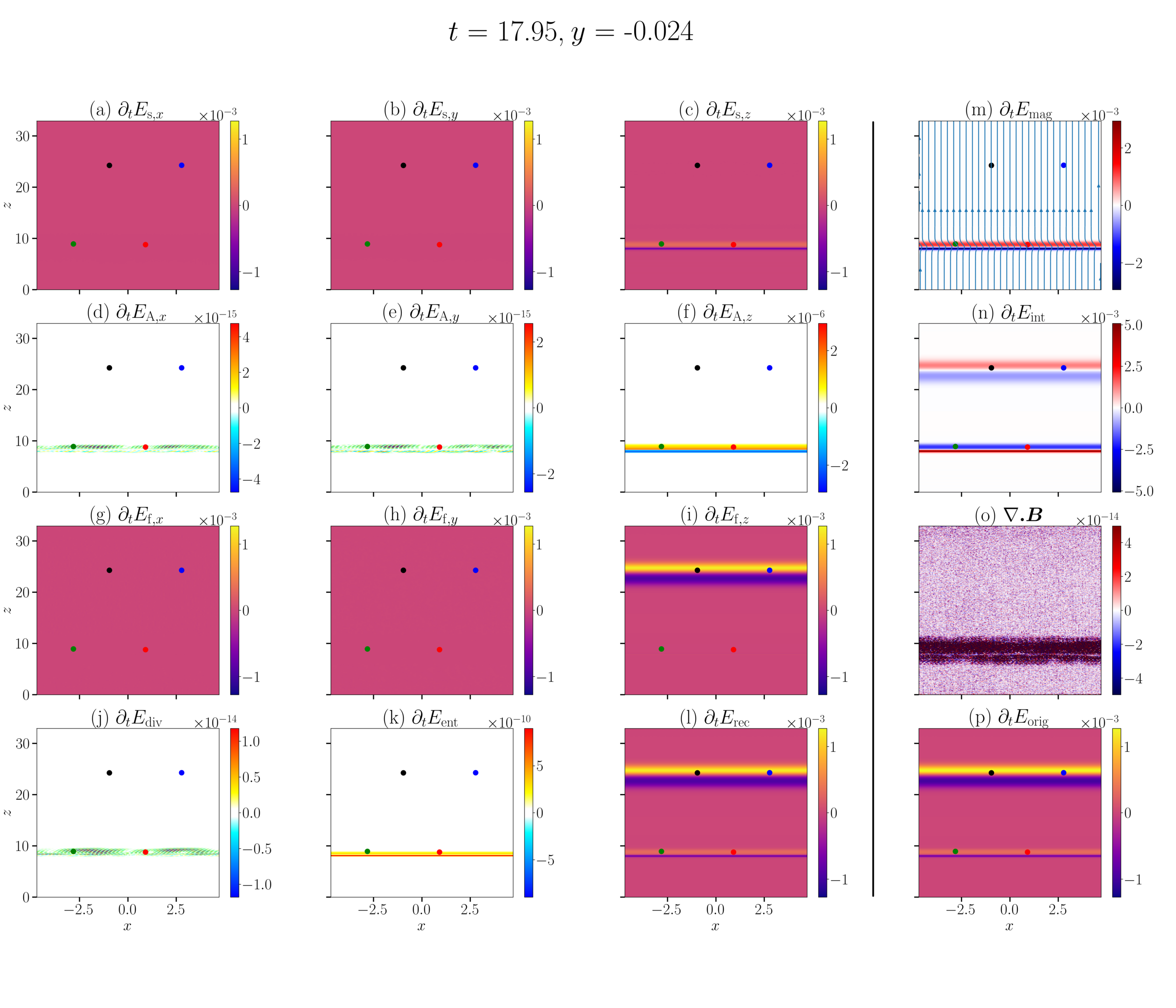}
    \caption{\textit{Left of the vertical separator line:} (a-l) $x$-$z$ plots of the eigenenergy time derivatives summed over both characteristic directions ($\pm$ terms) given by \autoref{eq:EDT} computed for simulation Ia ($\beta = 9.03$) at $t = 17.95$ and $y = -0.024$. The top nine panels (a-i) in a row-wise order from top to bottom are the $x$, $y$, and $z$ components of the (row 1) slow, (row 2) \alf{}, and (row 3) fast eigenenergy time derivatives, given by \autoref{eq:ESlowdt}, \ref{eq:EAlfdt}, and \ref{eq:EFastdt}, respectively. Panels (j) and (k) depict the field-divergence and entropy pseudo-mode eigenenergy time derivatives provided by \autoref{eq:EDivdt} and \ref{eq:EEntdt}, respectively, after summing over $q$. Panel (l) is the net rate of change of the total wave-energy recovered by summing over all the previous plots from (a) through to (k). 
    \textit{Right of the vertical separator line:} all quantities in panels m to p are computed directly from numerical differentiation of the primitive code output variables. (m, n) are the rates of change of the magnetic and internal (acoustic) energies, respectively (the fourth and third terms in \autoref{eq:ddtentot}). Panel (o) depicts the direct numerical results of $\div{\bs{B}}$. 
    Finally, panel (p) represents the original time derivative of the total wave-energy (\autoref{eq:EnTot}).
    The four coloured dots appearing in the plots are set up to coincide with the initial wavefront and move upwards at the three characteristic speeds, namely, the slow (red), \alf{} (green), fast (blue), plus the (noncharacteristic) sound (black) speed, provided for tracking the speed of the individual simulated wave branches post-decomposition. Given the different ranges of magnitude of the variables and in order to maximise the contrast between the waves and the background color for better visibility, as well as better categorization, we use three different color maps.
    Panels (d), (e), (f), (j), and (k) on the left of the vertical separator, as well as panels (m), (n), and (o) on the right of the separator have all been scaled according to their individual maxima and minima, using two different color maps. The rest of the plots are scaled with reference to panel (p).}
    \label{fig:highBetaDDTEnergy}
\end{figure}

\begin{figure}
    \centering    \includegraphics[width=\textwidth,height=\textheight,keepaspectratio]{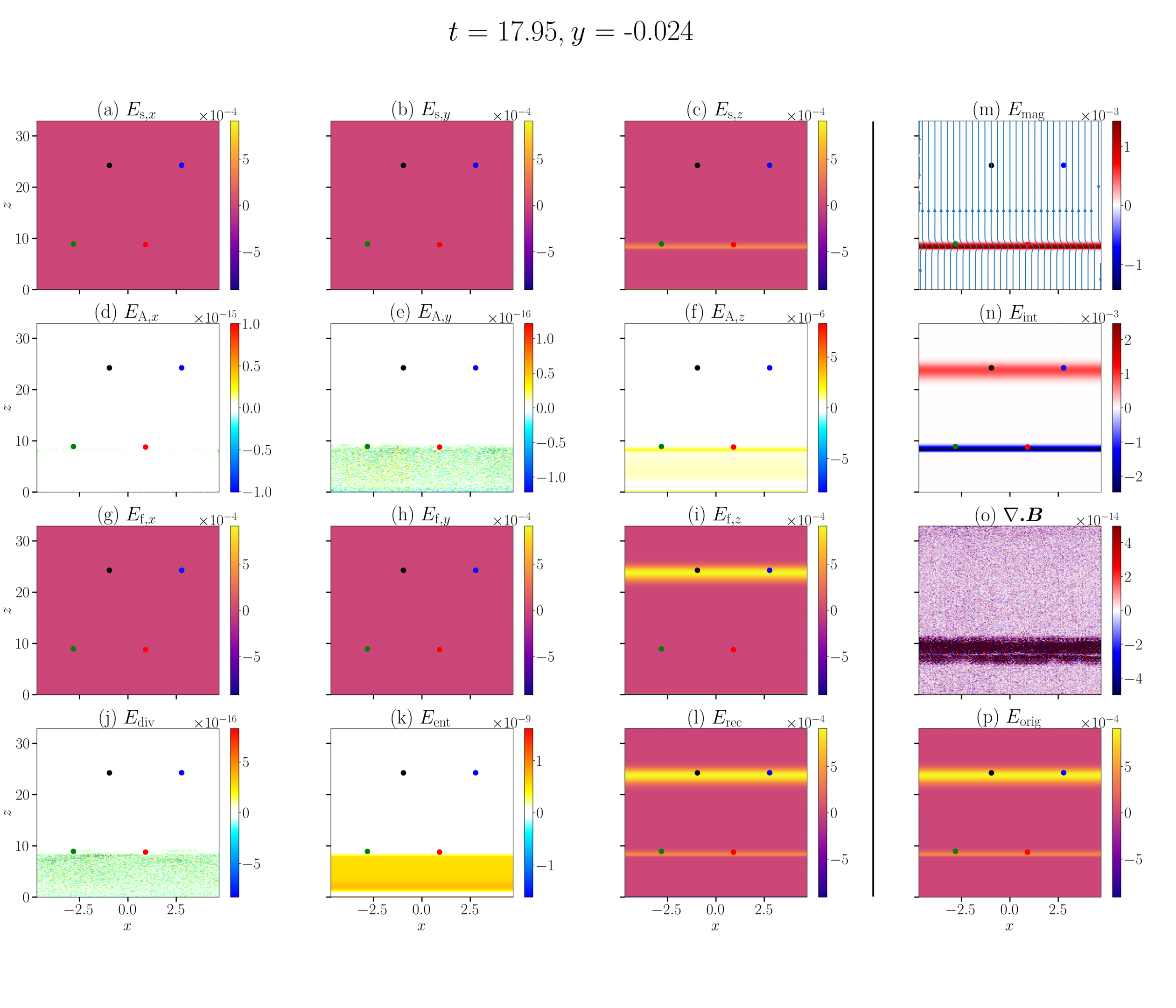}
    \caption{Plots of the energy contents corresponding to simulation Ia ($\beta = 9.03$) 
    taken at $y = -0.024$ and $t = 17.95$, given by the individual terms in \autoref{eq:totendecomposed} after summing over the characteristic directions ($\pm$), and presented using the same layout as that of \autoref{fig:highBetaDDTEnergy}. Notice the visible nonlinearity in the magnetic energy plot (m) manifested by the bends in the field lines at the location of the (magnetic) slow/\alf{} wavefront. The negative energy values in the internal energy (n) imply that acoustic energy is being extracted from the background state, meaning that the magnetically dominated (magneto\-acoustic) slow wavefront causes rarefaction in the gas (downtick in the thermal pressure) as it travels. This, however, is countered by larger positive values of magnetic energy (m), implying that magnetic energy is being deposited by the wave (uptick in the magnetic pressure), leading to an overall positive energy state for the slow branch. 
    Finally, note the extremely small (numerically zero) amounts of energy in the $x$ and $y$ components of the plots, as expected by the $x$- and $y$-independent nature of both the driver and the background magnetic field.
    }
    \label{fig:highBetaEnergy}
\end{figure}

\begin{figure}
    \centering    \includegraphics[width=\textwidth,height=\textheight,keepaspectratio]{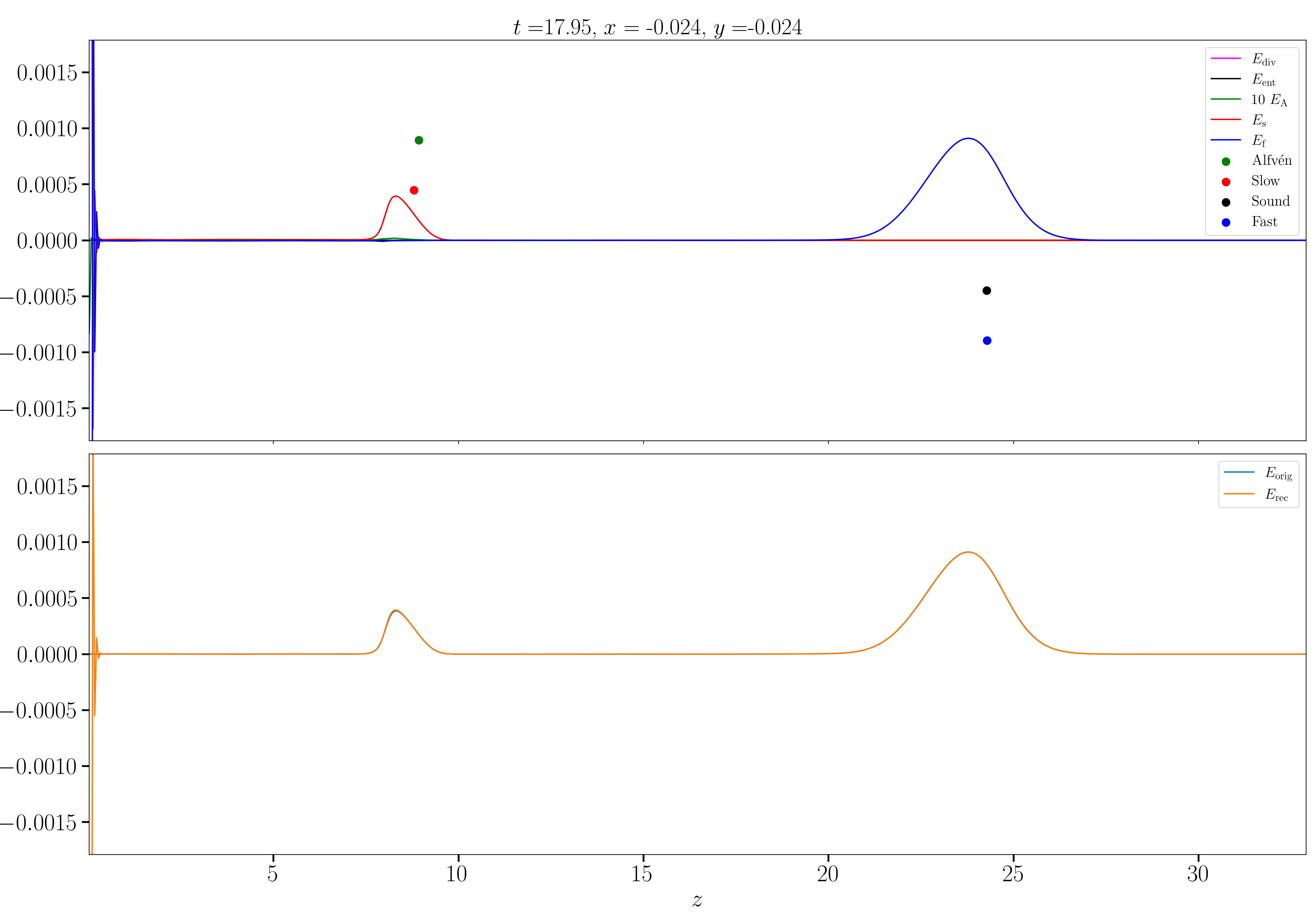}
    \caption{1D plots of wave-energy propagating along the $z$-axis belonging to simulation Ia ($\beta = 9.03$), taken at $x = y = -0.024$ and $t = 17.95$. \textit{Top:} Post-decomposition wave-energy components associated with the pseudo and physical modes given by \autoref{eq:totendecomposed}, individually summed over the $\pm$ terms and $q$. 
    \textit{Bottom:} The original and recovered energies of the wave.}
    \label{fig:highBetaEnergy1D}
\end{figure}

\begin{figure}
    \centering    \includegraphics[width=\textwidth,height=\textheight,keepaspectratio]{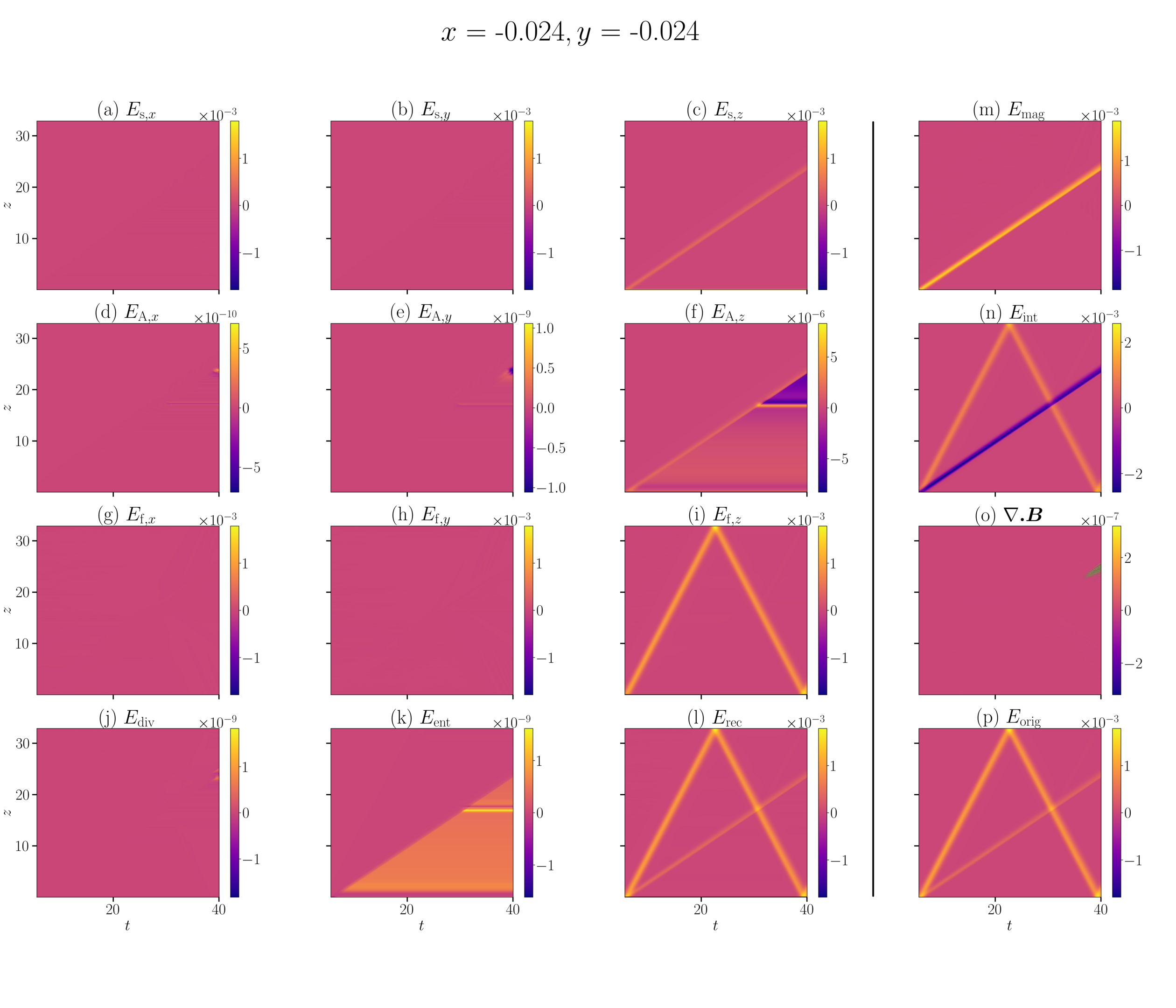}
    \caption{Time-distance ($t$-$z$) plots of the post-decomposition wave-energy components of simulation 1a ($\beta = 9.03$) at $x = y = -0.024$. The layout of the panels is carried over directly from \autoref{fig:highBetaDDTEnergy}.}
    \label{fig:highBetaPixel}
\end{figure}

The plots on the left of the vertical separator line in \autoref{fig:highBetaDDTEnergy} depict 2D $x$-$z$ slices of the individual eigenenergy time derivatives, summed over the characteristic directions ($\pm$ terms) in each of \autoref{eq:EDT}, shown in the plane $y = -0.024$ at $t = 17.95$. These are the three components of the (row 1: a-c) slow, (row 2: d-f) \alf, and (row 3: g-i) fast physical eigenenergy time derivatives, as well as the $q$-summed (j) field-divergence and (k) entropy pseudo-mode eigenenergy time derivatives, respectively, and (l) the net post-decomposition recovered rate of change of the wave-energy (computed by simply summing over panels (a) through to (k). The plots to the right of the separator all show quantities computed directly from the primitive output variables of the code: these are the rates of change of the (m) magnetic and (n) internal energies, (o) the field divergence $\div{\bs{B}}$, and lastly (p) the rate of change of the original wave-energy.  In summary, panels to the left of the vertical separator represent what the time-derivatives \emph{should be} given the local spatial derivatives (as per Equations \ref{eq:EDT}), while those to the right show the direct numerical time differences between two simulation outputs. The four colored points track the vertical distance traveled by different modes, calculated by integrating the $z$-component of their respective speeds over time, starting when the pulse is initiated.  These are the three characteristic modes, namely, the slow (red), \alf{} (green), and fast (blue) modes, plus a hypothetical sound wave (black), which can be compared with either the fast or slow mode in the appropriate $\beta$ limit.

Firstly, as seen in frame (p) belonging to the rate of change of the original wave-energy $\partial_t E_{\text{orig}}$, the initial pulse has split into two branches, with the leading branch (higher) moving at the fast speed and the trailing branch (lower) moving at the slow speed. This is perfectly recovered by the decomposition method, as shown in panel (l) corresponding to $\partial_t E_{\text{rec}}$, which indicates a high degree of numerical precision of the method. Moreover, observe how the two leading and trailing branches have been sifted out by the method and identified as the $z$ component of the fast (panel i) and slow (panel c) waves, respectively, moving at their corresponding characteristic speeds. In addition to the two dominant modes, some amount of \alf{} wave is also detected by the method, as seen in panel (f).

Note the visible nonlinearity of the system manifested in panel (m) by the skewed magnetic field lines at the location of the (magnetically dominated) slow and \alf{} waves. Also note the near equal wave speeds between the slow and \alf{} branches on the one hand, and the near equivalence of the fast branch and a pure hydrodynamic sound wave on the other hand, indicated by the approximate alignment of their respective dots. This further implies a near-degenerate state of the plasma between the slow and \alf{} waves on the one hand and the fast and sound waves on the other hand, due to the high value of $\beta$ (comparable to solar photospheric $\beta$ values). Nevertheless, despite the near-degeneracy, the modes will still be distinguishable based on the total variation of their state vector $\bs{P}$.

As a result, we expect that the (magnetically dominated) slow and \alf{}  branches, if extant, to coincide with one another, as well as with the location of any pure magnetic energy oscillations. This is clearly observed in panels (c), (f), and (m) of \autoref{fig:highBetaDDTEnergy}, belonging to the $z$ components of the slow and \alf{} eigenenergy time derivatives, and the rate of change of the magnetic energy, respectively. Furthermore, the additional coincidence between the lower pair of bands in panel (n), representing pure acoustic energy fluctuations, with those of the $z$ component of the slow branch implies the hybrid magneto\-acoustic nature of this wave, as one would expect.
On the other hand, the $z$ component of the (acoustically dominated) fast wave shown in panel (i) is seen to coincide only with the location of the upper pair of bands in panel (n), indicating the asymptotically acoustic nature of this wave. 
Moreover, as one would expect, based on the independence of the system on $x$ and $y$, there is virtually no flow of energy in these directions, as seen in panels (a), (b), (d), (e), (g), and (h). As for the pseudo-modes, it is more informative to inspect the actual energy profiles calculated by time-integrating the eigenenergy time derivatives, instead of the eigenenergy time derivatives \emph{themselves}.

\autoref{fig:highBetaEnergy} depicts the decomposed energy contents of the pulse as formulated in \autoref{eq:totendecomposed} for simulation Ia, in the same format as that of \autoref{fig:highBetaDDTEnergy} described above, and at the same location $y$ and time $t$. 
Note that while all the panels in the three left columns (a through to l) are the cumulative result of numerical time integration of their corresponding eigenenergy time derivatives, and hence contain information about the entire evolution of the system up to the timestamp, the panels in the rightmost column (m through to p) are directly computed from the single snapshot associated with the timestamp (with no integration involved).
The good correspondence between the two supports the validity of the decomposition.

Notice the numerically zero net energy associated with the field-divergence pseudo-mode $E_{\text{div}}$ shown in panel (j). This is due to the high fidelity of the simulation in maintaining the magnetic field divergence-free, as seen in panel (o). However, the method detects some small amounts of the net entropy pseudo-mode energy $E_{\text{ent}}$, as seen in frame (k), which is the result of a deviation from the assumed adiabaticity. The reason for this deviation is the mentioned nonlinearity of the system and its concomitant shock formation, leading to some numerical heating. Nevertheless, $E_{\text{ent}}$ is five orders of magnitude smaller than the total wave-energy $E_{\text{orig}}$ (or $E_{\text{rec}}$), meaning that the  departure from an adiabatic trajectory is in fact insignificant.

As noted before, the initial energy of the wave packet splits into the $z$ component of the slow (c), fast (i), and only small amounts of \alf{} (f) energies. The comparatively low magnitude of the \alf{} energy, in spite of the initial setup targeting the maximization of \alf{} waves, is the direct result of the driver not conserving momentum along the $z$-axis (in the ghost cells), generating more magneto\-acoustic wave-energy (see \autoref{eq:momentumzNotConserved} and the discussion above that).  This is further rendered mathematically intractable by the nonlinearity of the simulation, resulting in unpredictably complex behaviour of the solutions. Moreover, the coincidence between the bands in panels (f) and (k) implies some `leakage' of energy between the \alf{} energy and the entropy pseudo-energy, leading to the standing-wave trail in the \alf{} energy profile seen in frame (f). Nevertheless, the leading yellow band in panel (f), which coincides with the only band in panel (c), represents an accurately physical local decomposition at this height, since it sits above the band in panel (k) (compare the bands with reference to the green/red dot).

\autoref{fig:highBetaEnergy1D} shows a 1D cut of the decomposed energy contents (top) and the original and recovered energies (bottom) associated with the pulse along the $z$-axis, taken at $x = y = -0.024$ and $t = 17.95$. The \alf{} branch (green) is scaled up by a factor of 10 for visibility. Note the high numerical precision of the decomposition method in recovering the original wave-energy at all heights, except for some large error at the bottom boundary $z = 0$, and a minuscule mismatch at the peak location of the trailing slow wave, as seen in the bottom frame. The high amplitude (jagged) features near the driving boundary stem from a mismatch between the ghost cells and the first domain cell, generating a current sheet and shock at the boundary (see \autoref{sec:AppendixC}).

In \autoref{fig:highBetaPixel} we present time-distance (distance along $z$) plots of the decomposed energy components along the same line as in \autoref{fig:highBetaEnergy1D} (at $x = y = -0.024$), and following the same layout from Figure~\ref{fig:highBetaDDTEnergy}. The slope of the beams are determined by the propagation speeds of each branch. As seen in panels (c), (f), and (i) associated with the $z$ component of the slow, \alf, and fast waves, the slopes match the corresponding waves as identified by the decomposition scheme---near equal slopes for the slow and \alf{} waves and a steeper slope for the fast wave. Note the purple triangle in panel (f), with the vertex on the left marking the location of the nonlinear interaction of the upward propagating slow wave and the downward propagating fast wave (after having reflected from the top boundary), resulting in a standing \alf{} wave, with its ends anchored in the travelling slow and fast waves. We stress that this interaction could not have been detected with linear decomposition methods. The left vertex mentioned above is also clearly seen to be the point where the slow and fast waves meet at around $t \sim 30$, as shown in panel (l) or (p). 

As a final comment on these results, we note that both the value of $\div{\B}$ calculated using the derivative operators from the EEDM and the corresponding pseudo-energy $E_{\rm div}$ remain negligible throughout the simulation. We will see later that this is not the case in a fully 3D simulation, for reasons explained in more detail in \autoref{sec:AppendixE}.

\subsubsection{Simulation Ib: low-\texorpdfstring{$\beta$}{beta}}

\begin{figure}
    \centering    \includegraphics[width=\textwidth,height=\textheight,keepaspectratio]{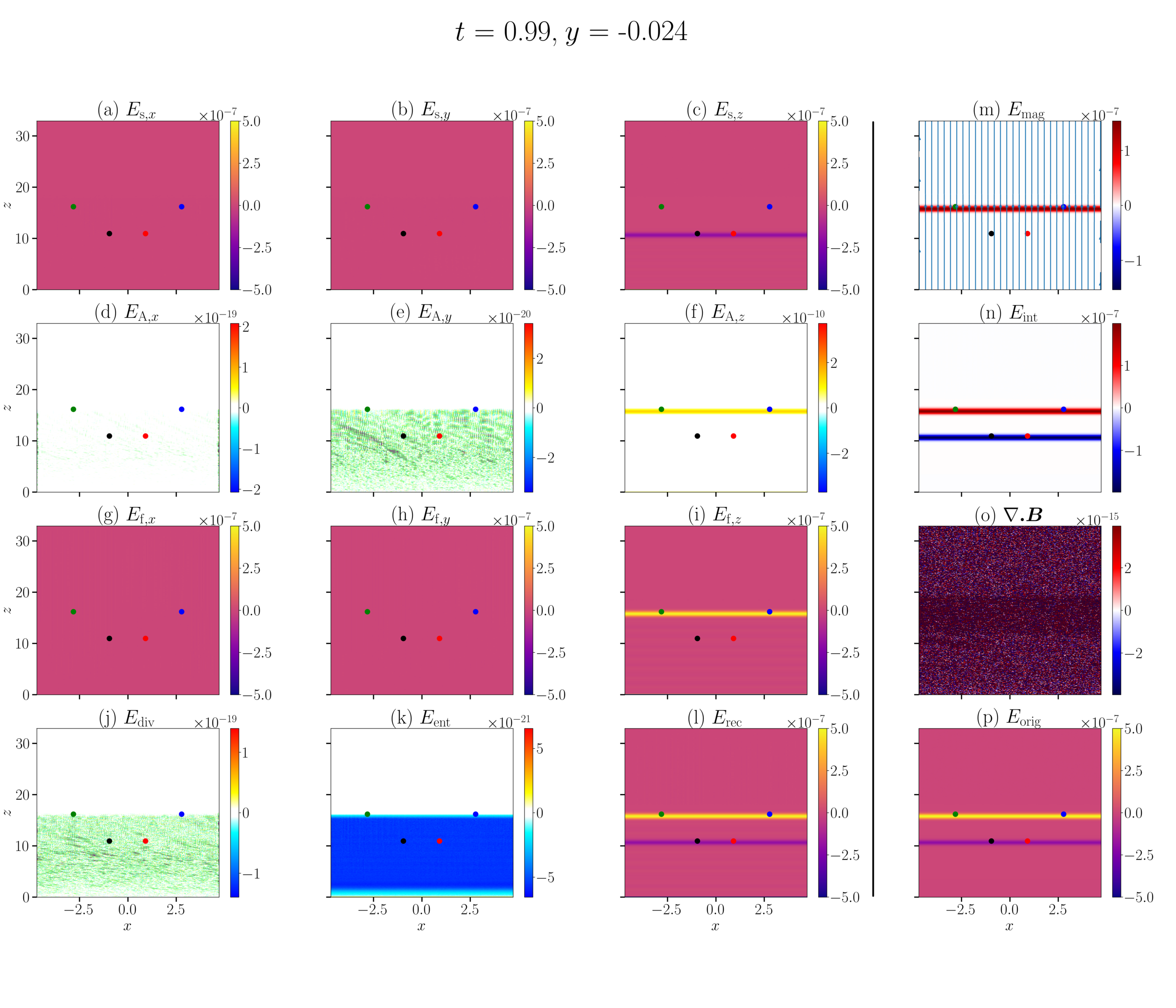}
    \caption{2D plots of the post-decomposition wave-energy components of simulation Ib ($\beta = 0.54$) at $y = -0.024$ and $t = 0.99$, presented in the same format as that of \autoref{fig:highBetaDDTEnergy}.}
    \label{fig:lowBetaEnergy}
\end{figure}

\begin{figure}
    \centering    \includegraphics[width=\textwidth,height=\textheight,keepaspectratio]{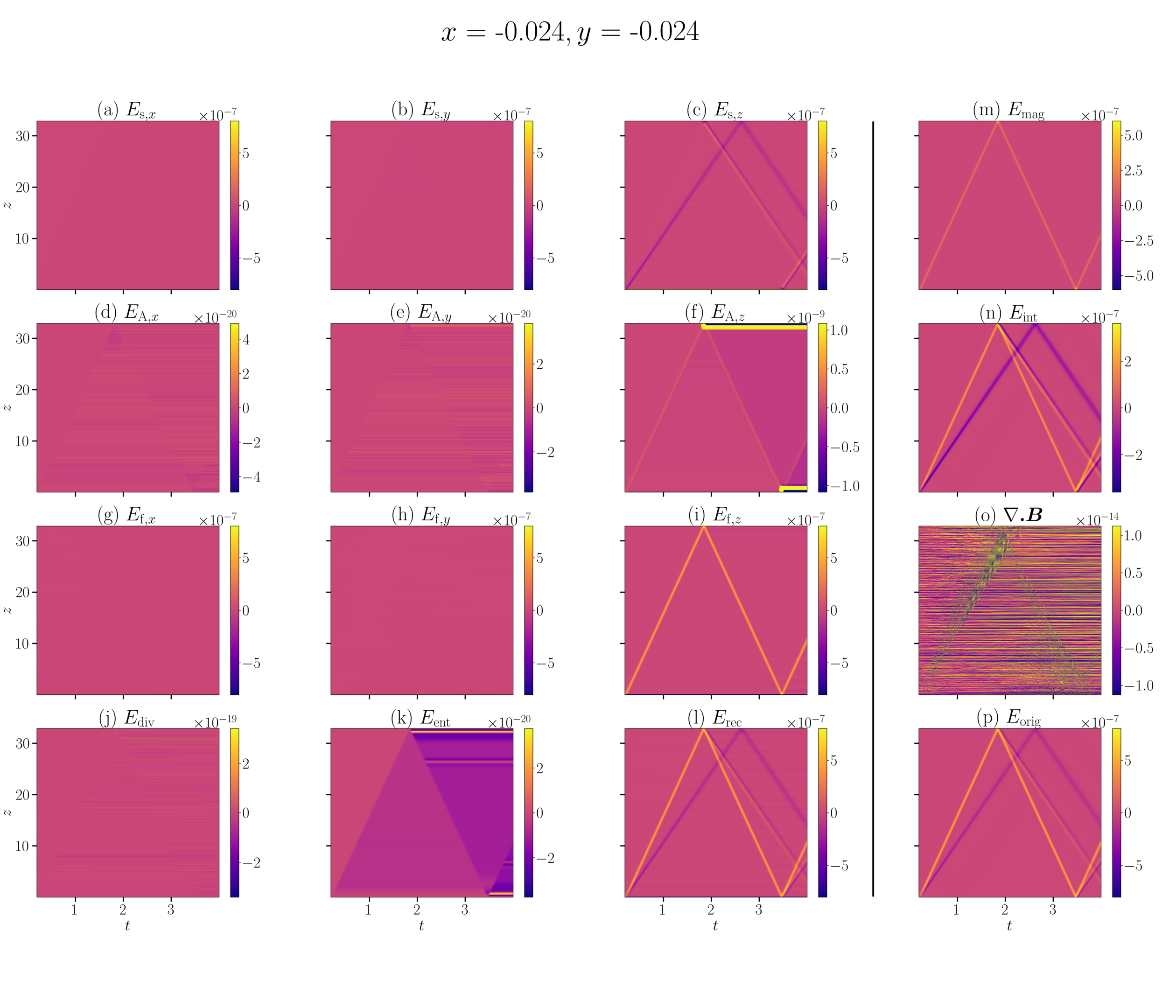}
    \caption{Time-distance ($t$-$z$) plots of the post-decomposition wave-energy components of simulation Ib ($\beta = 0.54$) at $x = y = -0.024$. The layout of the panels is carried over directly from \autoref{fig:highBetaDDTEnergy}.}
    \label{fig:lowBetaPixel}
\end{figure}
\autoref{fig:lowBetaEnergy} depicts the decomposed wave-energy contents associated with simulation Ib ($\beta = 0.54$) at $y = -0.024$ and $t = 0.99$. The ordering of the frames follows directly from \autoref{fig:highBetaDDTEnergy}. Here in a low-$\beta$ regime, the essence of the slow and fast magneto\-acoustic waves are the opposite of the previous simulation (Ia)---slow waves take on an asymptotically acoustic nature, while fast waves become magnetically dominated.  This is exactly captured by the decomposition method, judging by the coincidence of the wave-energy profiles in panels (f), (i), and (m) belonging respectively to $E_{\text{A},z}$, $E_{\text{f},z}$, and $E_{\text{mag}}$ on the one hand, and panels (c) and (n) corresponding to $E_{\text{s},z}$ and (the blue band of) $E_{\text{int}}$ on the other hand. Moreover, the near-alignment of the green (\alf) and blue (fast) dots, as well as the black (sound) and red (slow) dots, and their appropriate placement on the corresponding wave-energy branch, further corroborates the above interpretation.

Note that here, the slow wave has an overall negative amount of energy relative to the background, meaning that it extracts energy from the background state as it travels. This is due to a local diminution in the background gas pressure state, as evidenced by the lower band in panel (n), that is not compensated by a concomitant rise in the magnetic pressure. As a result, the (positive) input wave-energy provided by the driver in this case is carried predominantly by the (magnetic) fast wave, and only minimally by the \alf{} wave.

As a final remark on \autoref{fig:lowBetaEnergy}, notice the very minor amounts of the pseudo-mode energies in panels (j) and (k). However, while $E_{\text{div}}$ appears to be patternless and noise-like behind the wavefront, $E_{\text{ent}}$, similar to simulation Ia, has a defined shape that extends from the driver to the location of the fast/\alf{} wave. The latter observation indicates nonlinearity of the system and its associated numerical heating, which has led to some non-adiabaticity (caused by the generation of small disturbances in the polytropic coefficient $\kappa$).

\autoref{fig:lowBetaPixel} contains the time-distance plots of the various decomposed energy components of the wave-field of simulation Ib ($\beta = 0.54$), taken at $x = y = -0.024$ plotted over the entire simulation time. Note the equal slopes between panels (f), (i), and (m), respectively belonging to $E_{\text{A},z}$, $E_{\text{f},z}$, and $E_{\text{mag}}$, and panels (c) and parts of (n) associated with $E_{\text{s},z}$ and $E_{\text{ent}}$, indicating the equal propagation velocities explained above, as one would expect considering the low-$\beta$ state of the plasma and its associated near-degeneracy between the fast and \alf{} branches. Also note the horizontal beams at the top and bottom of panel (f) belonging to $E_{\text{A},z}$, indicating rather stronger standing \alf{} waves formed at the top and bottom boundaries, as a product of non-physical boundary conditions. Related are the two oblique parallel pairs of yellow and purple beams in panel (c) belonging to the $z$ component of the (acoustically dominated) slow wave, also caused by the combined interaction of the \alf{} and fast mode with the (admittedly unphysical) boundary conditions at $t\approx1.8$.
In this case, the slow mode generated by the boundary interaction has much less energy than the actual reflected fast mode.
The actual reflected slow mode, on the other hand ($t\approx2.5$), does not appear to generate any other waves on interaction with the BCs.
Apparently, the boundary conditions do reflect \alf{}, fast, and slow energy back into the simulation, and appear to do so somewhat cleanly for the fast and slow modes, but leave a standing perturbation at the boundary for the \alf{} mode.  One could presumably use this information to develop new boundary conditions that behave in a more preferable way.  We make no such attempt here, but only mention it to highlight the diagnostic potential of the proposed decomposition scheme.

\section{Simulation II: 3D structured magnetic field with a null}\label{sec:sim3D}

Our second simulation setup to further substantiate the EEDM is adapted from \cite{Galsgaard2003}. Much like the previous section, in simulation II we aim to formulate an approximately torsional (or twisting) driver to excite propagating incompressive \alf ic waves, this time travelling through a structured magnetic field containing a null point. Here we mimic \cite{Galsgaard2003} by introducing a rotational driving velocity on the bottom domain boundary, with this rotation being centered on the footpoint of the null point's spine line. \cite{Galsgaard2003} presented linear solutions using the WKB approximation showing how pure, torsional Alfv\'en waves should follow the field lines and spread out across the fan surface (without crossing it), while magneto\-acoustic waves should ``wrap'' in towards the null, and may transmit  energy across the fan plane. Disturbances with such characteristics were identified in full MHD simulations in the same paper. Here we perform a very similar simulation and apply our decomposition method, to test whether it can allow us to unambiguously identify these two different components of the wave field in a nontrivially structured magnetic field.

The simulation domain extends over $-1 \leq x,y,z \leq 1$, and is discretized using $385^3$ grid points, providing spatial resolution of $\approx 0.005$. The simulation is run for 3.6 non-dimensional time units at an output cadence of $dt = 0.009$, which is around
113 times smaller than the \alf{} time-scale, providing sufficient time resolution.

\subsection{Initial conditions}
The MHS model atmosphere of simulation II is comprised of a uniform plasma, superimposed with a 3D structured potential magnetic field including a null at the center of the box at $x = y = 0$ and $z = 1$, given by
\begin{equation}\label{eq:galsgaardfield}
    \bs{B}_0 = B_0 \lrp{x, y, -2\lrp{z - 1}},
\end{equation}
where $B_0 = 0.1$. Note that the $z$ component of the \alf{} speed is zero across the fan at $z = 1$, indicating that pure \alf{} waves cannot transmit through this plane. 
The thermodynamics of the system are prescribed by a uniform density and temperature $\rho_0 = 25$ and $T_0 = 0.1$ (the pressure and internal energy then follow from the ideal gas law). This sets the initial sound speed at $c_0 \approx 0.4$. Furthermore, given the nature of the magnetic field, the model atmosphere will have only one equipartition contour enclosing the null point, where the sound and \alf{} speeds coincide.  
It has a radius of 0.106 in the $y = 0$ plane.
Inside this contour $\beta>1$ and is formally infinite at the null (numerically, $\beta_\text{max}\approx 1230$ at cell centers), while outside of it $\beta<1$, with a minimum of $\beta_{\text{min}} \approx 0.0125$.  The magnetic field increasingly dominates the dynamics for decreasing $\beta$.  Outside the equipartition layer, fast (slow) waves are magnetically (acoustically) dominated, and vice versa inside. The non-dimensionalization factors for the Lare simulation are given by $L_N = 150$ km, $\rho_N= 2.007 \times 10^{-11}$ kg.m$^{-3}$, $T_N = 10^6$ K.

\subsection{BCs and driver}

The BCs are set to fully symmetric, i.e., $\partial_q{v_q} = 0$, $\partial_q{B_q} = 0$, and $\partial_q{Q} = 0$, where $Q$ is any of the thermodynamical variables,
on all faces of the simulation box, except for the bottom boundary plane where we implement another version of the torsional \alf ic driver suitable for handling the now variable magnetic field.  To formulate the driver, this time we start off by prescribing a 2D incompressive flow velocity profile, which is a slightly modified version to that of \cite{Galsgaard2003}, given by
\begin{equation}\label{eq:galsgaardVelocityBCs}
    \bs{v} = v_0 \exp\lrp{-\frac{x^2 + y^2}{2 \sigma_R^2}} \exp\lrp{-\frac{t - t_0}{2 \sigma_{\text P}^2}} \lrp{y, -x, 0},
\end{equation}
in which 
$t_0 = 0.581$ is the peak time of the \alf ic pulse,
$\sigma_R = 0.1$ sets the width of the pulse in the $xy$ plane,
$\sigma_{\text P} = 0.07$ (in combination with the \alf{} speed) sets the width of the propagating pulse along the $z$-axis, and
$v_0 = 1$ is the amplitude of the driver. In a similar fashion as before, the desired BCs on the magnetic field are then set by 
\begin{equation}\label{eq:galsgaardFieldBCs}
    \bs{B} = \bs{B}_0 - \sqrt{\rho_0} \bs{v}.
\end{equation}
This ensures that the field is initially line-tied, and incompressive perturbations are introduced in keeping with the velocity field according to the WKB solutions of \cite{Turkmani2004}. Finally, the BCs on the thermodynamics of the system are fixed to their respective initial values. 

It is worth emphasising that these BCs do not strictly satisfy the full nonlinear MHD equations, which will result in the type of boundary layer common in MHD simulations with driven boundaries. This has a (relatively minor) impact on how we interpret the energy decomposition results, described further below, and in \autoref{sec:AppendixE}.

\subsection{Results}

\begin{figure}
    \centering    \includegraphics[width=\textwidth,height=\textheight,keepaspectratio]{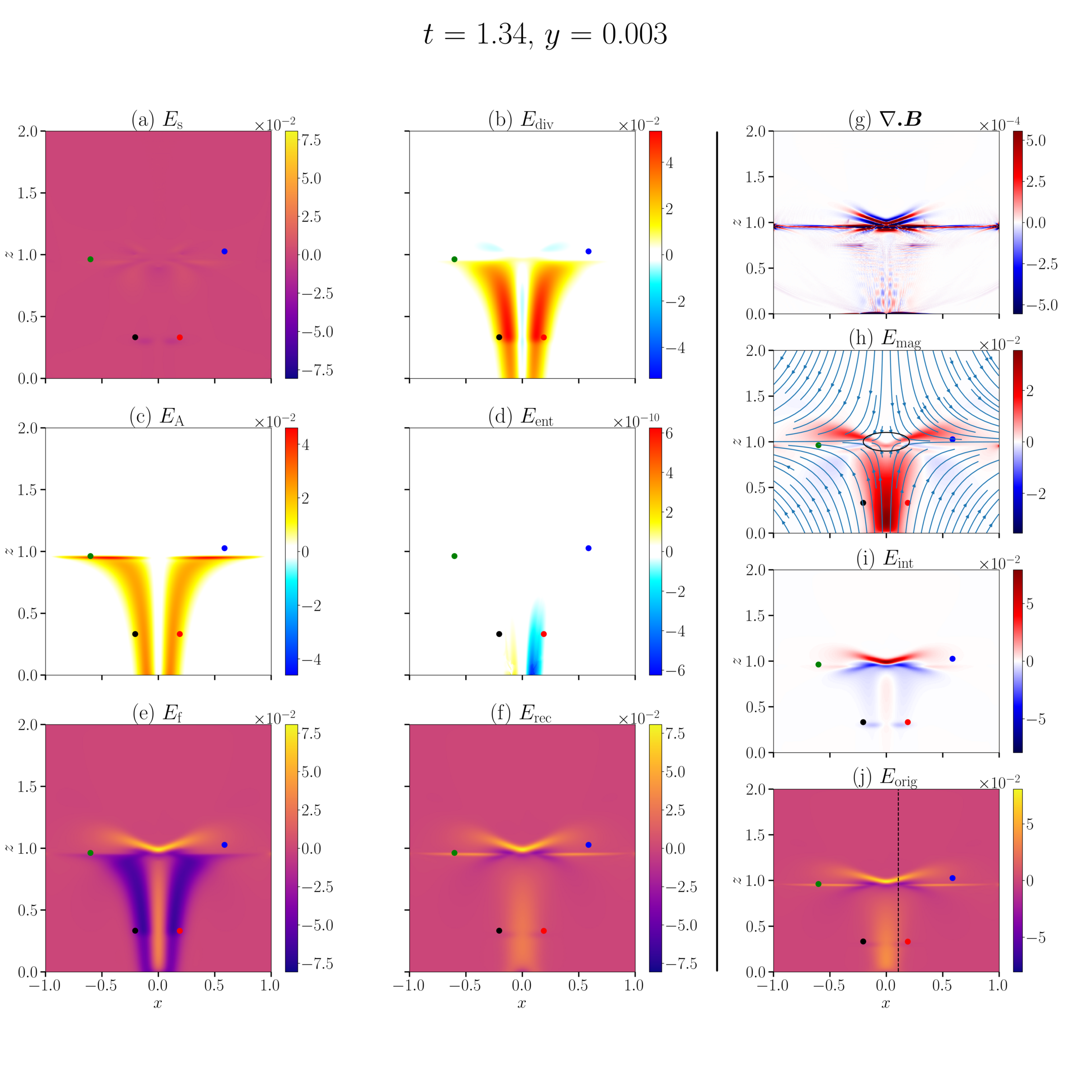}
    \caption{Plots of the decomposed energy components corresponding to simulation II taken at $y = 0.003$ and $t = 1.34$, given by the individual terms in \autoref{eq:totendecomposed}, summed over $q$ and both characteristic directions. \emph{Left of the vertical separator line:} The net amounts of energy of the (a) slow, (b) field-divergence, (c) \alf, (d) entropy, (e) fast modes and pseudo-modes, followed by (f) the recovered energy. \emph{Right of the vertical separator line:} Same as \autoref{fig:highBetaEnergy}. The vertical dashed line in panel (j) marks the $x$ location where we slice the plot range shown in this figure one more time to generate the 1D plots of \autoref{fig:galsgaardnx212ny192nt178Energy1D}, and 2D plots of \autoref{fig:galsgaard192PixelTotnx212}.}
    \label{fig:galsgaard192EnergyTot}
\end{figure}

\begin{figure}
    \centering    \includegraphics[width=\textwidth,height=\textheight,keepaspectratio]{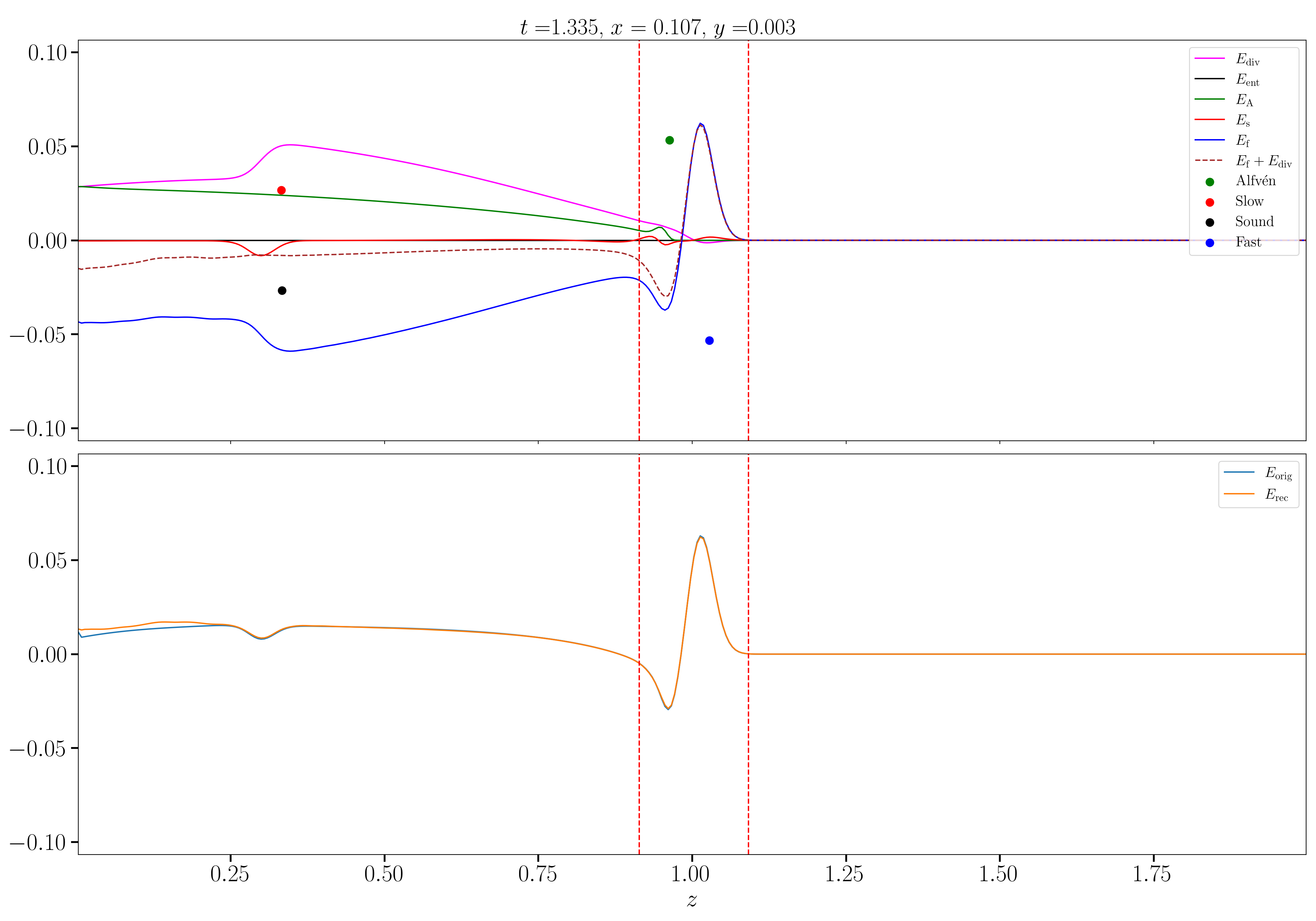}
    \caption{1D plots of wave-energy along the $z$-axis associated with simulation II, taken at $x = 0.107$, $y = 0.003$, and $t = 1.335$. The general format of the plot is carried over from that of \autoref{fig:highBetaEnergy1D}, with the only added feature being the brown dashed curve plotting $E_{\text{f}} + E_{\text{div}}$.}
    \label{fig:galsgaardnx212ny192nt178Energy1D}
\end{figure}

\begin{figure}
    \centering    \includegraphics[width=\textwidth,height=\textheight,keepaspectratio]{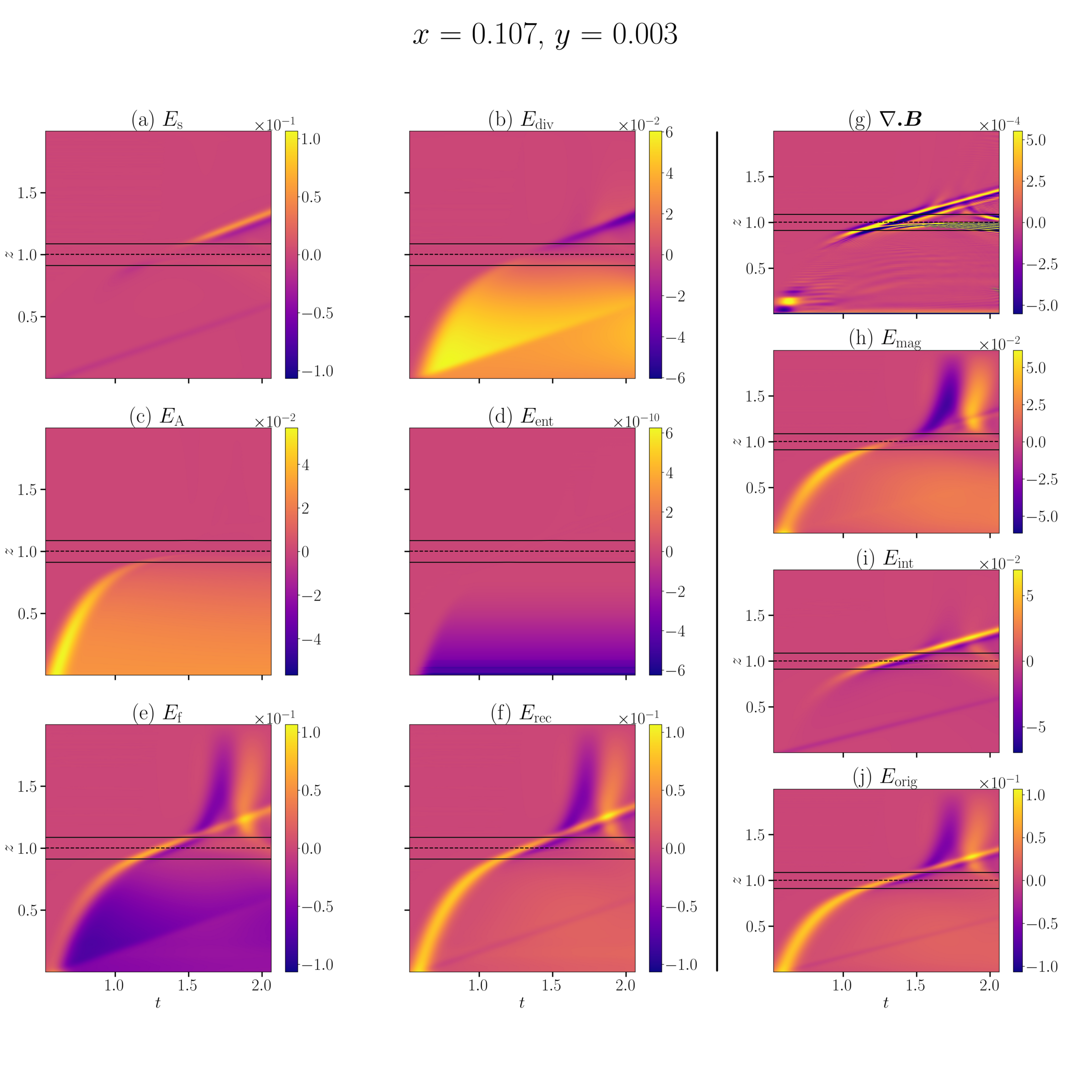}
    \caption{Time-distance ($z$-$t$) plots of wave-energy at $x = 0.107$, $y = 0.003$, belonging to simulation II. The ordering of the panels is directly inherited from \autoref{fig:galsgaard192EnergyTot}. The horizontal dashed line marks the location of the null, while the solid lines represent the locations of the equipartition level.}
    \label{fig:galsgaard192PixelTotnx212}
\end{figure}

\autoref{fig:galsgaard192EnergyTot} depicts the directionally summed (over both spatial and characteristic directions) branches of the physical and pseudo energies of simulation II, again calculated by numerically integrating the eigenenergy time derivatives over time. 
In this figure, panels (a) through to (f) to the left of the vertical separator line exhibit the net amounts of the slow ($E_{\text{s}}$), field-divergence ($E_{\text{div}}$), \alf{} ($E_{\text{A}}$), entropy ($E_{\text{ent}}$), fast ($E_{\text{f}}$), and the recovered ($E_{\text{rec}}$) energies and pseudo-energies, respectively. The panels on the right of the separator represent the same quantities as before (as per \autoref{fig:highBetaEnergy}), and are calculated directly from the single simulation time step.
We chose this more compact $q$- and characteristic direction-summed representation of the results to capture the most important aspects of each of the branches, as there are cancellations between the different directional components of  each branch due to the azimuthal symmetry of the simulation. 
Note how all the energy branches lie entirely in the vicinity of the driver at the bottom boundary, implying that the rotational driver does not induce any flow of energy in the horizontal directions: the energy flux is predominantly upwards along the field lines. The only added feature in this figure is the vertical dashed line in panel (j), which marks the $x$-coordinate where we slice the data further to inspect the 1D behaviour of the energies, as shown in \autoref{fig:galsgaardnx212ny192nt178Energy1D}.

Contrary to simulations Ia and Ib, panel (c) of \autoref{fig:galsgaard192EnergyTot} shows significant amplitudes associated with the \alf{} energy. Note how the \alf{} energy spreads out across the fan plane at $z = 1$, but cannot transmit through.
This is the expected behaviour since $a_z = 0$ across this plane, meaning that it acts as a shield against the upward propagation of \alf{} waves. However, upon interacting with the equipartition contour surrounding the null, the \alf{} waves can partially convert to magneto\-acoustic waves which in turn find their way into the upper half above the fan (see the features above the fan in panels a and e). 
Furthermore, note how the two horn-like features in panel (b) are aligned with the field lines, as one would expect from pure \alf{} waves. The footpoints of these features are rooted in the vicinity of the maxima of the spatial Gaussian function in \autoref{eq:galsgaardVelocityBCs}, around the $x^2 + y^2 - \sigma_R^2 = 0$ circle.
In view of the incompressibility of \alf{} waves and judging by the zero \alf-energy in the white regions of this panel, it appears that the driver must have triggered some thermal (and magnetic) pressure gradients in the 
volume adjacent to the spine ($z$-axis). This means that some amounts of magneto\-acoustic energy are likely to have been generated and channeled upwards through the `tube' formed in the middle of the \alf{} energy features, which is correctly seen to be the case in panel (e) belonging to the fast energy.

In the low $\beta$ region (away from the equipartition contour) below the fan, the \alf{} and fast modes are near-degenerate, yet have distinct variations to the MHD state vector $\bs{P}$ (compare, e.g., the left and right eigenvectors for the \alf{} and fast modes as given in \autoref{sec:AppendixA}). The near-degeneracy is indicated by the similar group velocities (blue and green dots), as well as the coincidence of the horn-like features in panels (c) and (e), which further unveils that up to a certain height below the equipartition contour, the fast energy is also propagating along the field lines.
These two modes slightly alter the background state as they are introduced by the driver and propagate through the system. In the EEDM, these appear as a decrease in $E_\text{f}$ and increase in $E_\text{A}$ (both relative to the background) below the fan surface.  The EEDM accurately captures the properties of two simultaneously propagating wave systems, as given by the fact that the energy in the \alf{} mode cannot and does not cross the fan surface at $z=1$, while the fast mode can and, indeed, does.

As the pulse nears the fan, all the energy branches show some contribution to the disturbance that follows the field lines, spreading out against the fan plane from underneath. At the same time, as shown in panel (e), a substantial amount of the energy begins to wrap around the null and transmit through to the upper region \citep[as observed by, e.g.,][]{2004A&A...420.1129M,Galsgaard2003}. This fast mode is in part generated by the boundary driving, and in part by mode conversion at the equipartition contour shown in panel (h). This is exactly what was previously observed by \cite{Galsgaard2003}.  Examining panel (e) in more detail, the fast energy can be broken down into three parts: (i) the vertical positive-energy orange beam extending along the spine up to the null, (ii) the purple negative-energy features in the driven sheath, and (iii) the V-shaped structure above the null. As touched on before, (i) is the result of thermal pressure as well as magnetic pressure fluctuations (triggered, at least in part, by the induced vertical velocity described in Appendix \ref{sec:AppendixC}) inside the cavity confined by the \alf{} energy (in panel c).  Note that this feature is positive in value and does not coincide with any other features from other energy branches (except for the faint vertical blue beam in panel b, which is negligible), and therefore appears to be a pure fast mode confined near the spine.
Thus, it is at least partially responsible for the transmission of fast energy into the upper half in the form of the V-shape above the fan.   The purple features (ii), are negative in value, which simply means that this part of the fast wave is extracting energy from the background due to the nonlinear nature of the evolution changing the background state.  Notice how (ii) can be further broken down to the lower dimmer purple part extending from the bottom boundary up to $z \approx 0.25$, and the upper darker part. While the darker portion is the actual propagating pulse itself, the dimmer part marks its wake, where nonlinear background disturbances are excited by the passing of the pulse through the plasma. Such nonlinear disturbances lead to numerical heating, and hence generate the (very small) entropy pseudo-energy in panel (d).
Due to its coincidence and similarity (both in the dimmer and darker regions) with $E_{\text{div}}$ (b), it appears that the mentioned extracted fast energy is somehow linked to the observed field-divergence pseudo-energy. 
We speculate that perhaps the field-divergence pseudo-energy, which is the result of some numerical pitfalls (see \autoref{sec:AppendixE}), is essentially leaking from the fast energy.
We will have a closer look at this speculation shortly.
Lastly, (iii) is the net result of the transmission of (i), as well as mode conversions from $E_{\text{A}}$ and/or $E_{\text{div}}$. 
 
The wrapping of the fast mode around the null is mainly due to the variable fast speed radially away from the spine, where at any given height, it is at its minimum at the location of the spine ($x = y = 0$), and increases hyperbolically with $x$ and $y$. Consequently, the fast waves, either directly transmitted or converted from \alf{} waves, that are near the spine will refract in towards the null and criss-cross above it in higher energy concentrations forming the maximum of the fast energy right above the null (the sharp bright orange vertex of the V-shape), while the waves farther from the spine move at higher speed and refract in to locations higher above the fan (the wings of the V-shape). One can take on a WKB approach and think about the refraction pattern in terms of rays intersecting in higher densities right above the null and rarefying in the wings.

There is a rather minimal energy contribution from slow waves (panel a). 
A faint signal does propagate upward from the lower boundary at the background slow wave speed (black/red dot), providing further validation of the decomposition. In addition, there is a second set of features evident in $E_{\text{s}}$ that appear around the fan surface after the \alf{} and fast modes reach that height.  This is also seen in the red curve of \autoref{fig:galsgaardnx212ny192nt178Energy1D}, inside the equipartition contour.  In the time animations this second feature does not propagate all the way from the boundary---but rather appears following the arrival of the fast/\alf{} pulse in the vicinity of the equipartition surface---indicating that this feature in the slow mode energy is generated by mode conversion of the cospatial \alf{} and fast mode pulses.
These slow modes were not identified in \citet{Galsgaard2003}, but appear analogous to the converted slow modes exiting a null point identified in 2D by \citet{Tarr2017} and in 3D by \citet{thurgood2017}, after some effort; they are identified trivially by the present decomposition scheme.

One important caveat to the above results is illustrated in \autoref{fig:galsgaard192EnergyTot}(b), which shows that significant energy appears in the field-divergence pseudo-mode, and (to a much smaller extent) the entropy pseudo-mode in panel (d). We remind the reader that the very presence of any non-zero pseudo-energies ultimately boils down to the detection of non-adiabaticity (departure from the polytropic gas assumption) and/or a non-zero field divergence as computed by the EEDM. Thus, to correctly interpret the results, one must take heed in identifying any sources of differences between how the simulating code and the decomposition scheme \emph{numerically} evaluate these two quantities, i.e., how the spatial partial derivatives are calculated. This is expounded in detail in \autoref{sec:AppendixE}. But in short, in order to numerically construct the terms in \autoref{eq:EDT}, we have first interpolated all physical variables (i.e., the plasma state vector $\bs{{P}}$) to the same grid locations (due to the staggered stencil used by Lare), and then evaluated the required spatial derivatives using a cubic B-spline scheme. This means that there is a mismatch between the derivative values used in the decomposition scheme and those same derivatives represented in Lare's native numerical scheme. This is particularly illustrated by the non-zero value of $\div{\B}$ in \autoref{fig:galsgaard192EnergyTot}, which is in contrast to the native zero value (to machine precision) of $\div{\B}$ in Lare. We stress that the high-order differentiation methods were found to be required to accurately recover the original wave-energy (\autoref{fig:galsgaardnx212ny192nt178Energy1D}). However, this comes at the cost of a leaky partitioning of energy between the modes in certain circumstances that remain to be explored fully. 
This illustrates the care that should be taken in implementing the decomposition scheme and interpreting its results. Additionally, recall that the quantities shown to the left of the vertical separator in the figure are integrated over time, whereas those in the rightmost column are computed instantaneously for the given snapshot. This implies that even low amounts of $\div{\B}$ or non-adiabaticity in the decomposition scheme in individual snapshots can potentially generate high values of pseudo-mode energies, due the time integration accumulating effects from all the previous snapshots.

\autoref{fig:galsgaardnx212ny192nt178Energy1D} depicts 1D plots of the various energy-decomposed branches of simulation II along a line in $z$ at $x = 0.107$, $y = 0.003$, which is signified by the vertical dashed line in \autoref{fig:galsgaard192EnergyTot}(j), and $t = 1.335$. The layout of the panels follows from \autoref{fig:highBetaEnergy1D}. However, for more clarity and in view of the mentioned similarities between the $E_{\text{f}}$ and $E_{\text{div}}$ profiles, we have now added the dashed brown curve which represents $E_{\text{f}} + E_{\text{div}}$. Note the rather small mismatch between the original and recovered energies in the bottom panel over $0 \leq z \lessapprox 0.25$. This is discussed in detail in \autoref{sec:AppendixE}. But in short, it stems from discontinuities introduced by the driver (seen as the cusp point near $z = 0$ in the blue curve belonging to $E_{\text{orig}}$), owing to the lack of complete nonlinear ideal-MHD solutions discussed before. As a result, the energy allocations performed by the method are slightly off in this region. Nevertheless, the actual pulse is located above this height, where the original energy recovery is precise, making the decomposition scheme valid inside the propagating wave. 

Notice the near anti-symmetry in trend (not magnitude) of the blue ($E_{\text{f}}$) and the magenta ($E_{\text{div}}$) curves in the left side of the upper frame up to the first equipartition level. 
This is more clearly observed in the dashed brown 
curve plotting their net effect $E_{\text{f}}$ + $E_{\text{div}}$, which 
exhibits a similar trend to that of the \alf{} wave.
This provides further evidence on the speculation touched on above that the detected $E_{\text{div}}$ might have been leaked from the fast branch, and that had the field-divergence been numerically zero as measured by the EEDM, the actual fast energy branch would have been the brown dashed curve belonging to $E_{\text{f}}$ + $E_{\text{div}}$. However, further investigation is required to substantiate this conjecture. Overall, comparing the two panels from $z \approx 0.25$ up to the left vertical dashed line, we can see that the total positive propagating wave-energy supplied by the driver at the particular point $x = 0.107$ and $y = 0.003$ is mainly being carried by the \alf{} and slow modes. However, as the wave approaches the equipartition contour, mode conversion switches the wave-energy to be predominantly carried by the fast branch. Moreover, in the time evolution of this plot, it is seen in our animations that as the fast branch hits the upper equipartition level (right vertical dashed line), it seems to start propagating faster than the background fast speed (blue dot) at first glance. However, this is only due to fast rays from elsewhere having refracted and found their way into this location.

\autoref{fig:galsgaard192PixelTotnx212} shows the time-distance ($t$-$z$) plots of the $q$- and characteristic direction-summed energy components decomposed by the method, taken at the same location as marked by the vertical dashed line in \autoref{fig:galsgaard192EnergyTot}(j), i.e., at $x = 0.107$ and $y = 0.003$. We remind the reader that the local slope of the distinctive filamentary features are given by the projection of the local speed of the corresponding wave branch onto the spatial cut. For instance, notice the constant speed of propagation of the slow branch in panel (a), as indicated by the straight filaments at the top and bottom of the panel, due to the nearly constant slow wave speed along this line, except near the equipartition layer.  This is in turn due to the constant background sound speed on the one hand, and the acoustically dominated nature of the slow waves away from the null  on the other hand. 
In contrast, the fast mode speed increases away from the null along this cut, causing a telltale bending of the energy density curve in panel (e). The more smeared out features in all panels indicate refracted waves originally launched from other locations crossing paths with the vertical dashed line in \autoref{fig:galsgaard192EnergyTot}(j).
The \alf{} energy depicted in panel (c) is seen to be continually blocked by the fan plane at $z = 1$, as expected. The fast energy in panel (e) is comprised of (i) an upward propagating positive part seen as the orange filament, and (ii) the purple regions, which manifest the near anti-symmetry with $E_{\text{div}}$ (of panel b) discussed above. Finally, the discontinuous slow energy features above the equipartition contour in panel (a) imply that the resultant slow wave must have been produced via mode transformations from other branches.

\section{Conclusion}\label{sec:conc}

In this paper, we formulated an exact method in \autoref{sec:math} which we term the \emph{Eigenenergy Decomposition Method} (EEDM), for analysing general nonlinear  ideal-MHD disturbances by locally teasing out the energy contributions due to the slow, \alf, and fast components. As the only precondition in deriving the equations, we assumed adiabaticity, which follows directly from the familiar polytropic equation ($p = \kappa \rho^{\gamma}$ with constant $\kappa$).
The usefulness, practicality, and precision of the method was then tested and demonstrated through three nonlinear numerical simulations.
The mathematical method is predicated on the method of characteristics. Specifically, given the ideal MHD solutions, we may analytically extract transformation matrices that locally transform the rate of change of the energy flux onto the characteristics of the MHD equations everywhere. Thereby, it is possible to break down the energy density flux vector in terms of its various MHD modes and pseudo-modes. These are entirely encapsulated in \autoref{eq:EDT} and \autoref{eq:totendecomposed}. Analytically, these equations remain well-defined everywhere, except for cusp points in the solutions introduced by discontinuous shocks describing jumps in the evolution of the plasma. Numerically, as we saw in \autoref{sec:sim3D}, cusp points can also stem from plugging in incomplete solutions as boundary conditions leading to discontinuous jumps in the evolution of the solutions, in addition to physical shocks.
Either way, we may still numerically evaluate the EEDM system of equations, bearing in mind that the energy partitioning may contain inaccuracies detectable by comparing the original and recovered energies. Such inaccuracies would potentially be small for smooth shocks.

Section \ref{sec:sim1D} was devoted to analysing two nonlinear MHD simulations involving initially-uniform plasmas of high and low plasma-$\beta$, respectively, threaded by an initially-vertical background field. These were driven by horizontally uniform torsional motions at the bottom boundary plane. The results unequivocally demonstrate that the decomposition method captures all of the expected properties of MHD waves: (i) each branch of energy is observed to propagate at the correct speed; (ii) the near-degenerate slow (fast) wave is correctly detected to be magnetically (acoustically) dominated in high-$\beta$ (simulation Ia), and vice-versa in low-$\beta$ (simulation Ib); (iii) the coincidence of the purely magnetic \alf{} energy as well as magnetically dominated magneto\-acoustic branches with the magnetic energy is clearly witnessed, and the same is true of the acoustically dominated branches and the internal energy; (iv) nonlinear wave-wave interactions and any resulting mode transformations (physical or artificial) are picked up. These are compelling indications that the method can indeed sift out the energy components of nonlinear mixed-mode disturbances.

In \autoref{sec:sim3D}, we presented the analysis of a more sophisticated 3D simulation of a uniform low-$\beta$ plasma superimposed by a structured magnetic field containing a null point, devised based on the study of \cite{Galsgaard2003}. Again, all of the expected features of the waves mentioned above were recovered by the method. Moreover, the eigenenergy decomposed results showed accurate general agreement with the findings of the above mentioned paper. However, in their assessment of the wave types, \cite{Galsgaard2003} classified the waves generated at the driving boundary to be entirely ``helical \alf'' waves, while our results illustrate that they are a mix of \alf{} and (magnetically dominated) fast waves, in addition to a weak slow-mode contribution. 

In each of the simulations, mode conversions not predicted by linear wave theory, and in some cases not identified in previous simulations, are revealed by the EEDM. For example, the nonlinear wave-wave interactions in simulations I, and the slow-mode waves exiting the null in simulation II, of which the latter were not identified in the \citet{Galsgaard2003} study.
This highlights the decomposition method as a promising tool for future analysis of dynamic MHD evolutions involving mixed-mode wavefields.

In spite of the above successes, some purely numerical challenges (independent of the decomposition method), expounded on in detail in \autoref{sec:AppendixE}, gave rise to the detection of pseudo-energies, which required a more careful examination of the results. Briefly, these challenges ensued from (i) lack of exact nonlinear MHD solutions to perfectly set up the driver, inevitable (ii) numerical heating causing departures from ideal MHD and (iii) differences between the numerical differentiation schemes used in the simulation code (Lare3d) and the decomposition method. Based on our observations, we suspect that using a higher order accurate simulation code operating on a uniform stencil will resolve the numerical challenges to a very good extent. However further work is needed to investigate this matter in more depth.

Finally, this is the first of a series of papers to be presented in the future, aiming to first explore the numerical challenge associated with the pseudo-energy generation in more detail, in addition to incorporating other physics in the EEDM such as  gravity and the Hall effect.

\section*{Acknowledgement}
Many thanks to Prof. Paul Cally, for his helpful comments and insightful discussions.
This work was performed on the OzSTAR national facility at Swinburne University of Technology. The OzSTAR program receives funding in part from the Astronomy National Collaborative Research Infrastructure Strategy (NCRIS) allocation provided by the Australian Government, and from the Victorian Higher Education State Investment Fund (VHESIF) provided by the Victorian Government.
Computations were run on the Australian National Computational Infrastructure's Gadi machine through an award from Astronomy Australia Ltd.'s Astronomy Supercomputer Time Allocation Committee.
L.A.T. is supported by the National Solar Observatory.

\software{Analytic work performed in Wolfram Mathematica \citep{Mathematica}. Simulations performed in Lare 3.4.1 \citep{ARBER2001151}. Numerical analysis performed in NumPy \citep{Harris:2020} and SciPy \citep{Virtanen2020}, and figures were prepared using the Matplotlib library \citep{Hunter:2007}, all in Python3.}

%




\appendix

\section{Left and right eigenvectors}\label{sec:AppendixA}

Here, we present the explicit expressions for the right ($R_q$) and left ($L_q$) eigenmatrices introduced in \autoref{sec:2p1}. 
The columns of the right eigenmatrices and rows of the left eigenmatrices, in order, correspond to the divergence, entropy, reverse \alf{}, forward \alf{}, reverse slow, forward slow, reverse fast, and forward fast eigenmodes, respectively.
The three components of the right eigenmatrix are given by
\begin{subequations}
    \begin{equation}
        R_x(\bs{x},t) = \left(
\begin{array}{cccccccc}
 0 & c^2 \rho  & 0 & 0 & \rho  & \rho  & \rho  & \rho  \\
 0 & 0 & 0 & 0 & -c_{\text{s},x} & c_{\text{s},x} & -c_{\text{f},x} & c_{\text{f},x} \\
 0 & 0 & -\frac{a_z c^2 \rho }{a_y \sqrt{\mu_0 \rho }} & \frac{a_z c^2 \rho }{a_y \sqrt{\mu_0 \rho }} & -\frac{a_x a_y c_{\text{s},x}}{a_x^2-c_{\text{s},x}^2} & \frac{a_x a_y c_{\text{s},x}}{a_x^2-c_{\text{s},x}^2} & -\frac{a_x a_y c_{\text{f},x}}{a_x^2-c_{\text{f},x}^2} & \frac{a_x a_y c_{\text{f},x}}{a_x^2-c_{\text{f},x}^2} \\
 0 & 0 & \frac{c^2 \rho }{\sqrt{\mu_0 \rho }} & -\frac{c^2 \rho }{\sqrt{\mu_0 \rho }} & -\frac{a_x a_z c_{\text{s},x}}{a_x^2-c_{\text{s},x}^2} & \frac{a_x a_z c_{\text{s},x}}{a_x^2-c_{\text{s},x}^2} & -\frac{a_x a_z c_{\text{f},x}}{a_x^2-c_{\text{f},x}^2} & \frac{a_x a_z c_{\text{f},x}}{a_x^2-c_{\text{f},x}^2} \\
 c^2 \rho  & 0 & 0 & 0 & 0 & 0 & 0 & 0 \\
 0 & 0 & -\frac{a_z c^2 \rho }{a_y} & -\frac{a_z c^2 \rho }{a_y} & \frac{a_y c_{\text{s},x}^2 \sqrt{\mu_0 \rho }}{c_{\text{s},x}^2-a_x^2} & \frac{a_y c_{\text{s},x}^2 \sqrt{\mu_0 \rho }}{c_{\text{s},x}^2-a_x^2} & \frac{a_y c_{\text{f},x}^2 \sqrt{\mu_0 \rho }}{c_{\text{f},x}^2-a_x^2} & \frac{a_y c_{\text{f},x}^2 \sqrt{\mu_0 \rho }}{c_{\text{f},x}^2-a_x^2} \\
 0 & 0 & c^2 \rho  & c^2 \rho  & \frac{a_z c_{\text{s},x}^2 \sqrt{\mu_0 \rho }}{c_{\text{s},x}^2-a_x^2} & \frac{a_z c_{\text{s},x}^2 \sqrt{\mu_0 \rho }}{c_{\text{s},x}^2-a_x^2} & \frac{a_z c_{\text{f},x}^2 \sqrt{\mu_0 \rho }}{c_{\text{f},x}^2-a_x^2} & \frac{a_z c_{\text{f},x}^2 \sqrt{\mu_0 \rho }}{c_{\text{f},x}^2-a_x^2} \\
 0 & 0 & 0 & 0 & c^2 \rho  & c^2 \rho  & c^2 \rho  & c^2 \rho  \\
\end{array}
\right)
    \end{equation}

    \begin{equation}
        R_y(\bs{x},t) = \left(
\begin{array}{cccccccc}
 0 & c^2 \rho  & 0 & 0 & \rho  & \rho  & \rho  & \rho  \\
 0 & 0 & -\frac{a_z c^2 \rho }{a_x \sqrt{\mu_0 \rho }} & \frac{a_z c^2 \rho }{a_x \sqrt{\mu_0 \rho }} & -\frac{a_x a_y c_{\text{s},y}}{a_y^2-c_{\text{s},y}^2} & \frac{a_x a_y c_{\text{s},y}}{a_y^2-c_{\text{s},y}^2} & -\frac{a_x a_y c_{\text{f},y}}{a_y^2-c_{\text{f},y}^2} & \frac{a_x a_y c_{\text{f},y}}{a_y^2-c_{\text{f},y}^2} \\
 0 & 0 & 0 & 0 & -c_{\text{s},y} & c_{\text{s},y} & -c_{\text{f},y} & c_{\text{f},y} \\
 0 & 0 & \frac{c^2 \rho }{\sqrt{\mu_0 \rho }} & -\frac{c^2 \rho }{\sqrt{\mu_0 \rho }} & -\frac{a_y a_z c_{\text{s},y}}{a_y^2-c_{\text{s},y}^2} & \frac{a_y a_z c_{\text{s},y}}{a_y^2-c_{\text{s},y}^2} & -\frac{a_y a_z c_{\text{f},y}}{a_y^2-c_{\text{f},y}^2} & \frac{a_y a_z c_{\text{f},y}}{a_y^2-c_{\text{f},y}^2} \\
 0 & 0 & -\frac{a_z c^2 \rho }{a_x} & -\frac{a_z c^2 \rho }{a_x} & \frac{a_x c_{\text{s},y}^2 \sqrt{\mu_0 \rho }}{c_{\text{s},y}^2-a_y^2} & \frac{a_x c_{\text{s},y}^2 \sqrt{\mu_0 \rho }}{c_{\text{s},y}^2-a_y^2} & \frac{a_x c_{\text{f},y}^2 \sqrt{\mu_0 \rho }}{c_{\text{f},y}^2-a_y^2} & \frac{a_x c_{\text{f},y}^2 \sqrt{\mu_0 \rho }}{c_{\text{f},y}^2-a_y^2} \\
 c^2 \rho  & 0 & 0 & 0 & 0 & 0 & 0 & 0 \\
 0 & 0 & c^2 \rho  & c^2 \rho  & \frac{a_z c_{\text{s},y}^2 \sqrt{\mu_0 \rho }}{c_{\text{s},y}^2-a_y^2} & \frac{a_z c_{\text{s},y}^2 \sqrt{\mu_0 \rho }}{c_{\text{s},y}^2-a_y^2} & \frac{a_z c_{\text{f},y}^2 \sqrt{\mu_0 \rho }}{c_{\text{f},y}^2-a_y^2} & \frac{a_z c_{\text{f},y}^2 \sqrt{\mu_0 \rho }}{c_{\text{f},y}^2-a_y^2} \\
 0 & 0 & 0 & 0 & c^2 \rho  & c^2 \rho  & c^2 \rho  & c^2 \rho  \\
\end{array}
\right)
    \end{equation}

    \begin{equation}
        R_z(\bs{x},t) = \left(
\begin{array}{cccccccc}
 0 & c^2 \rho  & 0 & 0 & \rho  & \rho  & \rho  & \rho  \\
 0 & 0 & -\frac{a_y c^2 \rho }{a_x \sqrt{\mu_0 \rho }} & \frac{a_y c^2 \rho }{a_x \sqrt{\mu_0 \rho }} & -\frac{a_x a_z c_{\text{s},z}}{a_z^2-c_{\text{s},z}^2} & \frac{a_x a_z c_{\text{s},z}}{a_z^2-c_{\text{s},z}^2} & -\frac{a_x a_z c_{\text{f},z}}{a_z^2-c_{\text{f},z}^2} & \frac{a_x a_z c_{\text{f},z}}{a_z^2-c_{\text{f},z}^2} \\
 0 & 0 & \frac{c^2 \rho }{\sqrt{\mu_0 \rho }} & -\frac{c^2 \rho }{\sqrt{\mu_0 \rho }} & -\frac{a_y a_z c_{\text{s},z}}{a_z^2-c_{\text{s},z}^2} & \frac{a_y a_z c_{\text{s},z}}{a_z^2-c_{\text{s},z}^2} & -\frac{a_y a_z c_{\text{f},z}}{a_z^2-c_{\text{f},z}^2} & \frac{a_y a_z c_{\text{f},z}}{a_z^2-c_{\text{f},z}^2} \\
 0 & 0 & 0 & 0 & -c_{\text{s},z} & c_{\text{s},z} & -c_{\text{f},z} & c_{\text{f},z} \\
 0 & 0 & -\frac{a_y c^2 \rho }{a_x} & -\frac{a_y c^2 \rho }{a_x} & \frac{a_x c_{\text{s},z}^2 \sqrt{\mu_0 \rho }}{c_{\text{s},z}^2-a_z^2} & \frac{a_x c_{\text{s},z}^2 \sqrt{\mu_0 \rho }}{c_{\text{s},z}^2-a_z^2} & \frac{a_x c_{\text{f},z}^2 \sqrt{\mu_0 \rho }}{c_{\text{f},z}^2-a_z^2} & \frac{a_x c_{\text{f},z}^2 \sqrt{\mu_0 \rho }}{c_{\text{f},z}^2-a_z^2} \\
 0 & 0 & c^2 \rho  & c^2 \rho  & \frac{a_y c_{\text{s},z}^2 \sqrt{\mu_0 \rho }}{c_{\text{s},z}^2-a_z^2} & \frac{a_y c_{\text{s},z}^2 \sqrt{\mu_0 \rho }}{c_{\text{s},z}^2-a_z^2} & \frac{a_y c_{\text{f},z}^2 \sqrt{\mu_0 \rho }}{c_{\text{f},z}^2-a_z^2} & \frac{a_y c_{\text{f},z}^2 \sqrt{\mu_0 \rho }}{c_{\text{f},z}^2-a_z^2} \\
 c^2 \rho  & 0 & 0 & 0 & 0 & 0 & 0 & 0 \\
 0 & 0 & 0 & 0 & c^2 \rho  & c^2 \rho  & c^2 \rho  & c^2 \rho  \\
\end{array}
\right)
    \end{equation}
\end{subequations}
The corresponding left eigenmatrices are then computed by taking the inverse of $R_q$ as follows
\begin{subequations}
    \begin{align}
        &L_x(\bs{x},t) = \nonumber \\ &\left(
\begin{array}{cccccccc}
 0 & 0 & 0 & 0 & \frac{1}{c^2 \rho } & 0 & 0 & 0 \\
 \frac{1}{c^2 \rho } & 0 & 0 & 0 & 0 & 0 & 0 & -\frac{1}{c^4 \rho } \\
 0 & 0 & -\frac{a_y a_z \sqrt{\mu_0}}{2 c^2 \sqrt{\rho } \left(a_y^2+a_z^2\right)} & \frac{a_y^2 \sqrt{\mu_0}}{2 c^2 \sqrt{\rho } \left(a_y^2+a_z^2\right)} & 0 & -\frac{a_y a_z}{2 c^2 \rho  \left(a_y^2+a_z^2\right)} & \frac{a_y^2}{2 c^2 \rho  \left(a_y^2+a_z^2\right)} & 0 \\
 0 & 0 & \frac{a_y a_z \sqrt{\mu_0}}{2 c^2 \sqrt{\rho } \left(a_y^2+a_z^2\right)} & -\frac{a_y^2 \sqrt{\mu_0}}{2 c^2 \sqrt{\rho } \left(a_y^2+a_z^2\right)} & 0 & -\frac{a_y a_z}{2 c^2 \rho  \left(a_y^2+a_z^2\right)} & \frac{a_y^2}{2 c^2 \rho  \left(a_y^2+a_z^2\right)} & 0 \\
 0 & \frac{c_{\text{s},x}^2-a_x^2}{2 c_{\text{s},x} \left(c_{\text{f},x}^2-c_{\text{s},x}^2\right)} & -\frac{a_x a_y}{2 c_{\text{s},x} \left(c_{\text{f},x}^2-c_{\text{s},x}^2\right)} & -\frac{a_x a_z}{2 c_{\text{s},x} \left(c_{\text{f},x}^2-c_{\text{s},x}^2\right)} & 0 & -\frac{a_y}{2 \left(c_{\text{f},x}^2-c_{\text{s},x}^2\right) \sqrt{\mu_0 \rho }} & -\frac{a_z}{2 \left(c_{\text{f},x}^2-c_{\text{s},x}^2\right) \sqrt{\mu_0 \rho }} & \frac{a^2-c^2+c_{\text{f},x}^2-c_{\text{s},x}^2}{4 c^2 \rho  \left(c_{\text{f},x}^2-c_{\text{s},x}^2\right)} \\
 0 & \frac{a_x^2-c_{\text{s},x}^2}{2 c_{\text{s},x} \left(c_{\text{f},x}^2-c_{\text{s},x}^2\right)} & \frac{a_x a_y}{2 c_{\text{s},x} \left(c_{\text{f},x}^2-c_{\text{s},x}^2\right)} & \frac{a_x a_z}{2 c_{\text{s},x} \left(c_{\text{f},x}^2-c_{\text{s},x}^2\right)} & 0 & -\frac{a_y}{2 \left(c_{\text{f},x}^2-c_{\text{s},x}^2\right) \sqrt{\mu_0 \rho }} & -\frac{a_z}{2 \left(c_{\text{f},x}^2-c_{\text{s},x}^2\right) \sqrt{\mu_0 \rho }} & \frac{a^2-c^2+c_{\text{f},x}^2-c_{\text{s},x}^2}{4 c^2 \rho  \left(c_{\text{f},x}^2-c_{\text{s},x}^2\right)} \\
 0 & \frac{a_x^2-c_{\text{f},x}^2}{2 c_{\text{f},x} \left(c_{\text{f},x}^2-c_{\text{s},x}^2\right)} & \frac{a_x a_y}{2 c_{\text{f},x} \left(c_{\text{f},x}^2-c_{\text{s},x}^2\right)} & \frac{a_x a_z}{2 c_{\text{f},x} \left(c_{\text{f},x}^2-c_{\text{s},x}^2\right)} & 0 & \frac{a_y}{2 \left(c_{\text{f},x}^2-c_{\text{s},x}^2\right) \sqrt{\mu_0 \rho }} & \frac{a_z}{2 \left(c_{\text{f},x}^2-c_{\text{s},x}^2\right) \sqrt{\mu_0 \rho }} & -\frac{a^2-c^2-c_{\text{f},x}^2+c_{\text{s},x}^2}{4 c^2 \rho  \left(c_{\text{f},x}^2-c_{\text{s},x}^2\right)} \\
 0 & \frac{c_{\text{f},x}^2-a_x^2}{2 c_{\text{f},x} \left(c_{\text{f},x}^2-c_{\text{s},x}^2\right)} & -\frac{a_x a_y}{2 c_{\text{f},x} \left(c_{\text{f},x}^2-c_{\text{s},x}^2\right)} & -\frac{a_x a_z}{2 c_{\text{f},x} \left(c_{\text{f},x}^2-c_{\text{s},x}^2\right)} & 0 & \frac{a_y}{2 \left(c_{\text{f},x}^2-c_{\text{s},x}^2\right) \sqrt{\mu_0 \rho }} & \frac{a_z}{2 \left(c_{\text{f},x}^2-c_{\text{s},x}^2\right) \sqrt{\mu_0 \rho }} & -\frac{a^2-c^2-c_{\text{f},x}^2+c_{\text{s},x}^2}{4 c^2 \rho  \left(c_{\text{f},x}^2-c_{\text{s},x}^2\right)} \\
\end{array}
\right)
    \end{align}    
    
    \begin{align}
        &L_y(\bs{x},t) = \nonumber \\ &\left(
\begin{array}{cccccccc}
 0 & 0 & 0 & 0 & 0 & \frac{1}{c^2 \rho } & 0 & 0 \\
 \frac{1}{c^2 \rho } & 0 & 0 & 0 & 0 & 0 & 0 & -\frac{1}{c^4 \rho } \\
 0 & -\frac{a_x a_z \sqrt{\mu_0}}{2 c^2 \sqrt{\rho } \left(a_x^2+a_z^2\right)} & 0 & \frac{a_x^2 \sqrt{\mu_0}}{2 c^2 \sqrt{\rho } \left(a_x^2+a_z^2\right)} & -\frac{a_x a_z}{2 c^2 \rho  \left(a_x^2+a_z^2\right)} & 0 & \frac{a_x^2}{2 c^2 \rho  \left(a_x^2+a_z^2\right)} & 0 \\
 0 & \frac{a_x a_z \sqrt{\mu_0}}{2 c^2 \sqrt{\rho } \left(a_x^2+a_z^2\right)} & 0 & -\frac{a_x^2 \sqrt{\mu_0}}{2 c^2 \sqrt{\rho } \left(a_x^2+a_z^2\right)} & -\frac{a_x a_z}{2 c^2 \rho  \left(a_x^2+a_z^2\right)} & 0 & \frac{a_x^2}{2 c^2 \rho  \left(a_x^2+a_z^2\right)} & 0 \\
 0 & \frac{a_x a_y}{2 c_{\text{s},y} \left(c_{\text{s},y}^2-c_{\text{f},y}^2\right)} & \frac{a_y^2-c_{\text{s},y}^2}{2 c_{\text{s},y} \left(c_{\text{s},y}^2-c_{\text{f},y}^2\right)} & \frac{a_y a_z}{2 c_{\text{s},y} \left(c_{\text{s},y}^2-c_{\text{f},y}^2\right)} & -\frac{a_x}{2 \left(c_{\text{f},y}^2-c_{\text{s},y}^2\right) \sqrt{\mu_0 \rho }} & 0 & -\frac{a_z}{2 \left(c_{\text{f},y}^2-c_{\text{s},y}^2\right) \sqrt{\mu_0 \rho }} & \frac{c_{\text{f},y}^2-c^2}{2 c^2 \rho  \left(c_{\text{f},y}^2-c_{\text{s},y}^2\right)} \\
 0 & \frac{a_x a_y}{2 c_{\text{s},y} \left(c_{\text{f},y}^2-c_{\text{s},y}^2\right)} & \frac{c_{\text{s},y}^2-a_y^2}{2 c_{\text{s},y} \left(c_{\text{s},y}^2-c_{\text{f},y}^2\right)} & \frac{a_y a_z}{2 c_{\text{s},y} \left(c_{\text{f},y}^2-c_{\text{s},y}^2\right)} & -\frac{a_x}{2 \left(c_{\text{f},y}^2-c_{\text{s},y}^2\right) \sqrt{\mu_0 \rho }} & 0 & -\frac{a_z}{2 \left(c_{\text{f},y}^2-c_{\text{s},y}^2\right) \sqrt{\mu_0 \rho }} & \frac{c_{\text{f},y}^2-c^2}{2 c^2 \rho  \left(c_{\text{f},y}^2-c_{\text{s},y}^2\right)} \\
 0 & \frac{a_x a_y}{2 c_{\text{f},y} \left(c_{\text{f},y}^2-c_{\text{s},y}^2\right)} & \frac{a_y^2-c_{\text{f},y}^2}{2 c_{\text{f},y} \left(c_{\text{f},y}^2-c_{\text{s},y}^2\right)} & \frac{a_y a_z}{2 c_{\text{f},y} \left(c_{\text{f},y}^2-c_{\text{s},y}^2\right)} & \frac{a_x}{2 \left(c_{\text{f},y}^2-c_{\text{s},y}^2\right) \sqrt{\mu_0 \rho }} & 0 & \frac{a_z}{2 \left(c_{\text{f},y}^2-c_{\text{s},y}^2\right) \sqrt{\mu_0 \rho }} & -\frac{c_{\text{s},y}^2-c^2}{2 c^2 \rho  \left(c_{\text{f},y}^2-c_{\text{s},y}^2\right)} \\
 0 & -\frac{a_x a_y}{2 c_{\text{f},y} \left(c_{\text{f},y}^2-c_{\text{s},y}^2\right)} & \frac{c_{\text{f},y}^2-a_y^2}{2 c_{\text{f},y} \left(c_{\text{f},y}^2-c_{\text{s},y}^2\right)} & -\frac{a_y a_z}{2 c_{\text{f},y} \left(c_{\text{f},y}^2-c_{\text{s},y}^2\right)} & \frac{a_x}{2 \left(c_{\text{f},y}^2-c_{\text{s},y}^2\right) \sqrt{\mu_0 \rho }} & 0 & \frac{a_z}{2 \left(c_{\text{f},y}^2-c_{\text{s},y}^2\right) \sqrt{\mu_0 \rho }} & -\frac{c_{\text{s},y}^2-c^2}{2 c^2 \rho  \left(c_{\text{f},y}^2-c_{\text{s},y}^2\right)} \\
\end{array}
\right)
    \end{align}    
    
    \begin{align}
        &L_z(\bs{x},t) = \nonumber \\ &\left(
\begin{array}{cccccccc}
 0 & 0 & 0 & 0 & 0 & 0 & \frac{1}{c^2 \rho } & 0 \\
 \frac{1}{c^2 \rho } & 0 & 0 & 0 & 0 & 0 & 0 & -\frac{1}{c^4 \rho } \\
 0 & -\frac{a_x a_y \sqrt{\mu_0}}{2 c^2 \sqrt{\rho } \left(a_x^2+a_y^2\right)} & \frac{a_x^2 \sqrt{\mu_0}}{2 c^2 \sqrt{\rho } \left(a_x^2+a_y^2\right)} & 0 & -\frac{a_x a_y}{2 c^2 \rho  \left(a_x^2+a_y^2\right)} & \frac{a_x^2}{2 c^2 \rho  \left(a_x^2+a_y^2\right)} & 0 & 0 \\
 0 & \frac{a_x a_y \sqrt{\mu_0}}{2 c^2 \sqrt{\rho } \left(a_x^2+a_y^2\right)} & -\frac{a_x^2 \sqrt{\mu_0}}{2 c^2 \sqrt{\rho } \left(a_x^2+a_y^2\right)} & 0 & -\frac{a_x a_y}{2 c^2 \rho  \left(a_x^2+a_y^2\right)} & \frac{a_x^2}{2 c^2 \rho  \left(a_x^2+a_y^2\right)} & 0 & 0 \\
 0 & \frac{a_x a_z}{2 \left(c_{\text{s},z}^3-c_{\text{f},z}^2 c_{\text{s},z}\right)} & \frac{a_y a_z}{2 \left(c_{\text{s},z}^3-c_{\text{f},z}^2 c_{\text{s},z}\right)} & \frac{a_z^2-c_{\text{s},z}^2}{2 \left(c_{\text{s},z}^3-c_{\text{f},z}^2 c_{\text{s},z}\right)} & -\frac{a_x}{2 \left(c_{\text{f},z}^2-c_{\text{s},z}^2\right) \sqrt{\mu_0 \rho }} & -\frac{a_y}{2 \left(c_{\text{f},z}^2-c_{\text{s},z}^2\right) \sqrt{\mu_0 \rho }} & 0 & \frac{2 c_{\text{f},z}^2-2 c^2}{4 c^2 c_{\text{f},z}^2 \rho -4 c^2 c_{\text{s},z}^2 \rho } \\
 0 & \frac{a_x a_z}{2 c_{\text{f},z}^2 c_{\text{s},z}-2 c_{\text{s},z}^3} & \frac{a_y a_z}{2 c_{\text{f},z}^2 c_{\text{s},z}-2 c_{\text{s},z}^3} & \frac{a_z^2-c_{\text{s},z}^2}{2 c_{\text{f},z}^2 c_{\text{s},z}-2 c_{\text{s},z}^3} & -\frac{a_x}{2 \left(c_{\text{f},z}^2-c_{\text{s},z}^2\right) \sqrt{\mu_0 \rho }} & -\frac{a_y}{2 \left(c_{\text{f},z}^2-c_{\text{s},z}^2\right) \sqrt{\mu_0 \rho }} & 0 & \frac{2 c_{\text{f},z}^2-2 c^2}{4 c^2 c_{\text{f},z}^2 \rho -4 c^2 c_{\text{s},z}^2 \rho } \\
 0 & \frac{a_x a_z}{2 c_{\text{f},z}^3-2 c_{\text{f},z} c_{\text{s},z}^2} & \frac{a_y a_z}{2 c_{\text{f},z}^3-2 c_{\text{f},z} c_{\text{s},z}^2} & \frac{a_z^2-c_{\text{f},z}^2}{2 c_{\text{f},z}^3-2 c_{\text{f},z} c_{\text{s},z}^2} & \frac{a_x}{2 \left(c_{\text{f},z}^2-c_{\text{s},z}^2\right) \sqrt{\mu_0 \rho }} & \frac{a_y}{2 \left(c_{\text{f},z}^2-c_{\text{s},z}^2\right) \sqrt{\mu_0 \rho }} & 0 & -\frac{2 c_{\text{s},z}^2-2 c^2}{4 c^2 c_{\text{f},z}^2 \rho -4 c^2 c_{\text{s},z}^2 \rho } \\
 0 & -\frac{a_x a_z}{2 c_{\text{f},z}^3-2 c_{\text{f},z} c_{\text{s},z}^2} & -\frac{a_y a_z}{2 c_{\text{f},z}^3-2 c_{\text{f},z} c_{\text{s},z}^2} & \frac{c_{\text{f},z}^2-a_z^2}{2 \left(c_{\text{f},z}^3-c_{\text{f},z} c_{\text{s},z}^2\right)} & \frac{a_x}{2 \left(c_{\text{f},z}^2-c_{\text{s},z}^2\right) \sqrt{\mu_0 \rho }} & \frac{a_y}{2 \left(c_{\text{f},z}^2-c_{\text{s},z}^2\right) \sqrt{\mu_0 \rho }} & 0 & -\frac{2 c_{\text{s},z}^2-2 c^2}{4 c^2 c_{\text{f},z}^2 \rho -4 c^2 c_{\text{s},z}^2 \rho } \\
\end{array}
\right)
    \end{align}
\end{subequations}

\section{First sanity check: pure magneto\-acoustic waves}\label{sec:AppendixB}

In the absence of exact 3D nonlinear solutions, as a first sanity check of the proposed decomposition scheme, consider the following case involving 2D magneto\-acoustic waves. 
Such a system is comprised of a 2D magnetic field (uniform or otherwise) fixed in the plane of, say, $xz$, and driven by some perturbations in the direction of $x$, generating magneto\-acoustic (a mixture of slow and fast) waves. It is well known that under these assumptions, any perturbations (linear or otherwise) in the plane of the magnetic field will remain in that plane. This is due to a lack of dependence on the out-of-plane direction $y$ to couple the equations, and hence \alf{} waves would be polarized along this axis. Therefore, setting $B_y = v_y = 0$ in \autoref{eq:EAlfdt}, we find
\begin{align}
    \partial_t{E_{\rm A}} &= \sum_{n = 3}^{4} \boldsymbol{c}_n\boldsymbol{.}\overline{\mathcal{L}}_n = 0,
\end{align}
and hence $E_{\rm A} = f_{\text{A}}(x,y,z)$, where $f_\text{A}$ is the background \alf{} energy. However, since we began with no background \alf{} waves, we find that $f_\text{A} = 0$, and hence  $E_{\rm A} = 0$, which is the expected result.

\section{Notes on the torsional Alfv\'enic driver}\label{sec:AppendixC}

To further inspect the nature of \autoref{eq:turkmani2004} and \ref{eq:travellingAlf:Bperp}, which do not constitute exact nonlinear ideal-MHD solutions, and elaborate on the source of any non-\alf{} waves it generates, we may substitute them in \autoref{eq:3DidealMHD}. Doing so will result in all of the MHD equations being satisfied, except for the $z$-component of the momentum \autoref{eq:3DidealMHD:momentum}, meaning that our driver does not conserve momentum along the $z$-axis, and instead yields
\begin{equation}\label{eq:momentumzNotConserved}
    \partial_t{v_z} + v_z \partial_z{v_z} = \frac{-\lrp{z - a_0 t - z_0}}{\rho_0 \sigma^2} \exp{\lrp{-\lrp{\frac{z - a_0 t  - z_0}{\sigma}}^2}},
\end{equation}
which is nonlinear and has no known analytical solutions. However, we may find a linear solution assuming $v_z \ll 1$, which is found to be
\begin{equation}
    v_z = -\frac{1}{2 a_0 \rho_0} \exp\lrp{-\frac{z^2 + \lrp{z - a_0 t - z_0}^2}{\sigma^2}} \lrp{\exp\lrp{\lrp{\frac{z}{\sigma}}^2} - \exp\lrp{\lrp{\frac{z - a_0 t - z_0}{\sigma}}^2}}.
\end{equation}

Thus, using \autoref{eq:turkmani2004} and \ref{eq:travellingAlf:Bperp} as the bottom BCs in our simulations Ia and Ib (in the ghost cells), this deficiency (non-conservation of the vertical component of the momentum equation) comes at the cost of the simulation code generating some $v_z$ in the first interior cell to balance out the equations. In other words, in order for the ideal-MHD equations to evolve self-consistently, there has to be a response function $v_{\text{res},z}$ which counters the effect of the $v_z$ that stems from the driver not being a perfectly exact solution. 
Moreover, this non-zero $v_{\text{res},z}$ will then feed back into the other equations in \autoref{eq:3DidealMHD}, resulting in a deviation of the propagating solutions from the hardwired BCs, and thereby exciting some amounts of magneto\-acoustic waves in addition to \alf{} waves. This is inevitable in the absence of exact propagating nonlinear solutions, but not unfavourable in our case, as our goal is to merely detect and measure the amount of energy carried by all MHD wave-types present in the system.

\section{Second sanity check: Monochromatic torsional Alfv\'en waves}\label{sec:AppendixD}

Taking $B_\perp$ in \autoref{eq:turkmani2004} as constant, and assuming a uniform plasma permeated with a constant vertical field as initial conditions, we arrive at pure monochromatic \alf{} wave solutions given by
\begin{equation}\label{eq:monochromaticAlf}
    \boldsymbol{B}(z,t) = B_\perp\lrp{\cos\lrp{k z - \omega t + \varphi},\sin\lrp{k z - \omega t + \varphi}, \frac{B_0}{B_\perp}},
\end{equation}
and \autoref{eq:vperp}. These solutions can be seen to satisfy \autoref{eq:3DidealMHD}, and lead to a constant total energy given by
\begin{align}
    E_{\rm tot} = \frac{B_0^2 + B_\perp^2}{2} + \frac{p_0}{\gamma - 1},
\end{align}
indicating a constant net energy transport, i.e., $\partial_t{E_{\rm tot}} = 0$, which describes a monochromatic \alf{} wave. This result is then straightforwardly reproduced using the decomposition method by evaluating \autoref{eq:ddtdecomposed1} using \autoref{eq:EDT} and \ref{eq:monochromaticAlf}.

\section{Numerical pitfalls}\label{sec:AppendixE}

Here we highlight two numerical issues that impact the analysis of our simulation results, and that are relevant to future numerical implementations of the eigenenergy decomposition method. The first of these relates to departures from ideal MHD in the simulations, and the second to the derivative stencil employed in calculating the decomposition.

\subsection{Departures from ideal MHD}

As described explicitly in \autoref{sec:math}, the decomposition method is based on the ideal MHD system. It is well established that MHD codes do not satisfy these equations in the strict sense: some numerical dissipation is always present, even if explicit resistivity and viscosity are set to zero. This induces a deviation from the assumed adiabaticity, and results in some amounts of the wave-energy being attributed to the entropy pseudo-energy in regions with substantial dissipation (numerical or otherwise). Nevertheless, the ability of the decomposition method to accurately identify the different component wave mode energies still relies on the precision with which the original wave-energy is recovered. In any case, away from such regions, where the evolution is close to ideal, the decomposition should remain physical and accurate.

As seen in \autoref{fig:galsgaardnx212ny192nt178Energy1D} there is an imperfect match between the original ($E_{\text{orig}}$) and the recovered ($E_{\text{rec}}$) energies. This we attribute to non-ideal behaviour close to the driving boundary -- common in many MHD simulations -- for the following reasons.
First, the driver is set up in the ghost cells at the bottom boundary, and despite the smooth introduction of the signal into the domain through the temporal and spatial Gaussian envelopes, it still creates a spatial discontinuity between the first ghost cell and first interior cell. Such behaviour is clearly manifested by the cusp point in the blue curve ($E_{\text{orig}}$) in the left-most cell in the bottom panel of \autoref{fig:galsgaardnx212ny192nt178Energy1D}. Note that the time evolutions of ${\bf v}$ and ${\bf B}$ imposed in the ghost cells
do not constitute exact ideal-MHD solutions (mass conservation, the $z$-components of the momentum and inductions equations, and the energy equation are identically satisfied, but not the other four equations). This means that a discontinuity (including, e.g., a small-scale current sheet) is present between the ghost cells and the internal cells. As a result, there will inevitably be some numerical dissipation in the vicinity of the boundary plane. From \autoref{fig:galsgaardnx212ny192nt178Energy1D} it is also clear that the numerical dissipation in the vicinity of the boundary is no longer affecting the recovery of the original wave-energy above $z\approx 0.25$. Note that a similar behaviour is observed in the vicinity of our (symmetric reflective) side boundaries.

In general, these non-ideal effects essentially produce changes to the background, represented by either $E_0$ (for a single-time analysis) or $f$ (for the time-integrated analysis).  This suggests putting some effort into better ascribing energy ``lost'' from each mode to that background, which would facilitate a more straightforward interpretation of the decomposition in regions with dissipation.

\subsection{Derivative stencils and the divergence pseudo-mode}
One key explicit assumption of the mathematical formulation of \autoref{sec:math} is a strictly divergence-free magnetic field, see \autoref{eq:3DidealMHD:induction}. All MHD codes have their own method of ensuring that $\div{\B}$ is either very small, or zero to machine precision.  The latter is achieved in Lare3d with a constrained transport scheme on a staggered grid, which essentially enforces the integral form of Gauss' law on each computational cell \citep[see][for a detailed discussion]{ARBER2001151}.  Numerically, $\div{\B}$ will only be zero when calculated according to Lare's numerical scheme, and if one explicitly calculates $\div{\B}$ using some other derivative operator, a non-zero result is generally expected. As such, if different derivative operators are used in the decomposition method compared to the MHD code, this can lead to energy being attributed to the field-divergence pseudo-mode.  

Beyond just the magnetic field, Lare uses a second-order finite volume scheme with a Lagrangian (half) step to compute spatial partial derivatives on a staggered grid and advance the MHD solutions.  Effectively, Lare calculates the update to each primitive variable in fundamentally separate ways, with the result that there is no uniquely defined differential operator that is applied uniformly to all variables.  The overall scheme is self-consistent for Lare, of course, which reproduces the expected evolution of standard MHD tests.

The decomposition scheme requires different (and more sophisticated) combinations of the state variables and their spatial derivatives than exist in the MHD equations themselves, and it is not immediately obvious how well these will be represented by Lare (or any other MHD code).
We found that directly implementing the native Lare grid and differentiation scheme into the decomposition method yielded a poor recovery of the original wave-energy (i.e., a poor match between the directly computed energy density and sum of the decomposed wave energy densities).
Instead, an acceptable level of total energy conservation with the EEDM required higher-order derivative operators and a consistently defined finite difference stencil.
To implement this, we first applied appropriate averaging to interpolate the Lare magnetic and velocity variables to cell centers (the Lare pressure and density are natively defined at cell centers).
Then, we used the cubic B-spline smoothing interpolation (splrep) and differentiation (splev) routines provided in the Scipy library to compute the required derivatives on the cell-center uniform grid.
This procedure yielded the high fidelity energy decompositions and reconstructions achieved in this work.

We speculate that the interpolations and derivatives defined using the spline functions closely match the equivalent Lare versions in almost every case, except for the magnetic field divergence, owing to Lare's constrained transport scheme.  We believe this explains much of the puzzling behavior in simulation II.

The rather significant field divergence {(\emph{as calculated by using the derivative operators employed in the decomposition scheme}) in simulation II is shown in panel (g) of \autoref{fig:galsgaard192EnergyTot}, and} gives rise to substantial amounts of $E_{\text{div}}$ in panel (b). {When comparing the two, it is important to recall that panel (g) shows an instantaneous evaluation of $\div{\B}$, whereas $E_{\text{div}}$ is essentially a cumulative effect (since it is constructed by time integrating $\partial_t E_{\text{div}}$). Equivalent plots for simulations I (Figures \ref{fig:highBetaEnergy} and \ref{fig:lowBetaEnergy}) do not exhibit the same behaviour.}

Another interesting point to note is that in simulation II the positive energy density in $E_{\text{div}}$ appears to have a counterpart negative fast energy density (compare panels (b) and (e) of \autoref{fig:galsgaard192EnergyTot}). This is even more apparent in the 1D plot of \autoref{fig:galsgaardnx212ny192nt178Energy1D}. In that figure we include a brown dashed curve representing $E_{\text{div}}+E_{\rm f}$, which may be a more accurate representation of $E_{\rm f}$. It is not clear why the $E_{\text{div}}$ should be produced at the expense of an opposite-sign counterpart in the fast energy density specifically, but we speculate that the fast mode present in simulation II contains magnetic variations proportional to $\div{\B}$ that cancel precisely under Lare's constrained transport scheme but not after the spline interpolation and differentiation in the decomposition scheme.

As one final comment on the pseudo-energies, the many orders-of-magnitude difference observed between $E_{\text{ent}}$ and $E_{\text{div}}$ in \autoref{sec:sim3D} is due to two reasons. First, our inspections of the primitive variables show that the nonlinearities seeding $E_{\text{ent}}$ are not very sharp, and hence do not require extreme smoothing via the shock viscosity. Second, the entropy pseudo-mode only concerns the thermodynamical variables, which are already cell-centered in the simulation code, and hence require no further interpolation. Therefore, their spatial derivatives, which are used to evaluate the entropy pseudo-energies, are in good agreement between the simulation code and the EEDM. On the other hand, the simulation code handles the magnetic field in a much more sophisticated way (see above discussion on how Lare defines the MHD variables on the staggered grid), which makes it less amenable to accurate and compatible incorporation in the EEDM. Thereby, the net effect of the initial interpolation of the field components on the cell-centers followed by the cubic B-spline spatial differentiation of them leads to a more significant disparity between the simulation code and the EEDM in evaluating $\div{\B}$, than the adiabatic precondition.

The above considerations illustrate that care must be taken both in the implementation and interpretation of the decomposition method. In general, it is desirable to minimise numerical dissipation, and to match the derivative and interpolation operators used in the decomposition scheme to those used in the code that produces the data. As discussed above, this was not possible for the Lare code due to non-conservation of the total energy in the EEDM when using the Lare derivatives. We speculate that these issues should be much reduced when analysing output from a code that uses (i) a co-spatial grid for all variables, and (ii) high-order spatial differencing. This remains to be explored in future work.


\bibliographystyle{aasjournal}



\end{document}